\documentclass[12pt,preprint]{aastex}

\begin{document}
\title{
Spatially Resolved Galaxy Star Formation and its Environmental Dependence I
}

\author{Niraj Welikala\altaffilmark{1}, Andrew J. Connolly\altaffilmark{2},
        Andrew M. Hopkins\altaffilmark{3}, Ryan Scranton\altaffilmark{1}, Alberto Conti\altaffilmark{4}
       }

\altaffiltext{1}{Department of Physics and Astronomy, University of Pittsburgh, 
3941, O'Hara Street, Pittsburgh, PA 15260, USA,
welikala@phyast.pitt.edu}
\altaffiltext{2}{Department of Astronomy, University of Washington, Box 351580, 
Seattle, WA 98195-1580, USA,  ajc@astro.washington.edu}
\altaffiltext{3}{School of Physics, University of Sydney, NSW 2006, Australia,
ahopkins@physics.usyd.edu.au}
\altaffiltext{4}{Space Telescope Science Institute, Baltimore, MD 21218, USA
}

\begin{abstract}
We use the photometric information contained in individual pixels of
44,964 ($0.019<z<0.125$ and $-23.5<M_r<-20.5$) galaxies in
the Fourth Data Release (DR4) of the Sloan Digital Sky Survey
to investigate the effects of environment on galaxy star formation (SF). We
use the pixel-z technique, which combines stellar population synthesis models
with photometric redshift template fitting on the scale of individual
pixels in galaxy images. Spectral energy distributions are constructed, sampling a wide range of
properties such as age, star formation rate (SFR), dust obscuration and
metallicity. By summing the SFRs in the pixels, we demonstrate that the
distribution of total galaxy SFR shifts to lower values as the local density
of surrounding galaxies increases, as found in other studies. The effect
is most prominent in the galaxies with the highest star formation, and we
see the break in the SFR-density relation  at a local galaxy density of $\approx 0.05\,$(Mpc/h)$^{-3}$.
Since our method allows us to spatially resolve the SF distribution
within galaxies, we can calculate the mean SFR of each galaxy as
a function of radius. We find that on average the mean SFR is dominated
by SF in the central regions of galaxies, and that the trend for suppression
of SFR in high density environments is driven by a reduction in this nuclear
SF. We also find that the mean SFR in the outskirts is largely independent
of environmental effects. This trend in the mean SFR is shared by galaxies
which are highly star forming, while those which are weakly star forming
show no statistically significant correlation between their environment
and the mean SFR at any radius.
\end{abstract}


\section{Introduction
\label{sec:introduction}}

\subsection{Environmental dependences of galaxy properties and dependence on morphology
\label{subsec:intro1}}

According to models of hierarchical formation \citep{Kau:93,Som:99,Col:00}, galaxies form in less dense 
environments and are then accreted into larger halos (e.g., falling into
clusters or groups), having their hot gas reservoir removed as this occurs.
This predicts that galaxy properties, such as color, luminosity and SFR, are correlated with their environment. Galaxies in highly dense regions are predicted to be more luminous and more red than those in lesser dense regions. In fact, the mean galaxy environment as a function of color and luminosity has been explored using the Sloan Digital Sky Survey (SDSS) \citep{Hog:03,Bla:03}, as was the dependence of the color-magnitude relation of bulge-dominated galaxies on their environment \citep{Hog:04}. The models also predict a gradual decrease in star formation (SF) activity in more dense regions, as seen in recent observational studies exploring the variation of
star formation rate (SFR) with environment \citep{Lew:02,Gom:03}. Numerous
physical mechanisms for how such SFR suppression might occur have been
suggested, including ram pressure stripping of gas \citep{GG:72}, gravitational
interactions between galaxies \citep{BV:90} and between galaxies and
a non-uniform cluster potential \citep[``galaxy harassment," ][]{Moo:99}.

Recent N-body simulations and semi-analytical models have extended our
insight into a number of the important issues with the hierarchical
formation scenario. \citet{Bal:00}, investigating the origin
of cluster-centric gradients in SFRs and colors of rich cluster galaxies,
used a model where clusters are built through the ongoing accretion of
field galaxies. These models assume that after galaxies enter the
cluster their SFRs decline on a timescale of a few Gyr
which is the typical gas consumption timescale for disk galaxies in
the field. They combined these timescales with mass accretion histories
from N-body simulations of cluster formation in a $\Lambda$CDM universe to
show that there is an expected strong suppression of SF in
cluster galaxies. The simulations also show that a significant fraction of
galaxies beyond the virial radius of the cluster may have been within
the main body of the cluster in the past. This would explain why star
formation in the outskirts of clusters (and as far out as two virial
radii) is systematically suppressed relative to the field. The agreement
with the data beyond the cluster virial radius is further improved by
assuming that gas-stripping happens within lower mass systems,
before the galaxy is actually accreted into the cluster. The suggestion
is that the SFRs of cluster galaxies depend primarily on the the time
since their accretion onto the cluster, and that the SFR suppression
happens gradually over a few Gyr.

Here we investigate the relation between environment and SFR for galaxies in
the low redshift universe. \citet{Gom:03}, using H$\alpha$ equivalent widths
(EW) as an indicator of SFR in the Early Data Release (EDR) of the SDSS,
established an SFR-density relation for the SDSS, confirming a similar result
seen in the 2 degree Field Galaxy Redshift Survey (2dFGRS)
by \citet{Lew:02}. They found that the overall distribution of SFRs is
shifted toward lower values in more dense environments. The effect is most
noticeable for the strongly star-forming galaxies (EW(H$\alpha$)$ > 5\,$\AA)
in the 75th percentile of the SFR distribution. They also found a
characteristic ``break''  (or characteristic density) in the density-SFR
relation at a local galaxy density of $\approx 1\,$(Mpc/h)$^{-2}$.
\citet{Gom:03} explored whether the density-morphology relation \citep{Dre:80}
alone could explain the density-SFR relation, and concluded that it could not.
Using the concentration index of SDSS galaxies as a morphology indicator,
they showed that SFRs for galaxies of the same type were suppressed
in dense regions. Higher redshift samples ($z>0.2$) have also been used to
suggest a suppression in the SFR of galaxies in the cores of distant clusters compared to those in the field \citep{Bal:97,Has:98,Pog:99,Cou:01,Pos:01}. Together this provides strong evidence for a decrease in SFR of galaxies in dense environments, spanning a wide range of densities ($0.08-10\,$(Mpc/h)$^{-2}$) and redshift (out to $z\approx 0.5$). In addition, \citet{Bal:97} and \citet{Has:98} found that cluster galaxies have a reduced SFR compared with the field, independent of morphology.

Not all studies agree with the general SFR-density trend though. \citet{Bal:04}
 analyzed H$\alpha$ emission strength as a function of galaxy environment using galaxies selected from the SDSS and 2dFGRS. They
found that the distribution of H$\alpha$ EWs is bimodal, consisting of
actively star-forming populations with EW(H$\alpha$)$>4$\,\AA\ and a
quiescent population with little current SF. They showed that
the distribution of EW(H$\alpha$) for the star forming population does
not itself depend on environment, and concluded that it was unlikely that SFRs
are gradually decreasing in a substantial number of star-forming galaxies in
or near dense regions today. They did find, however, that the fraction of
galaxies with EW(H$\alpha$)$>4\,$\AA\ decreases steadily with increasing
local density.

\subsection{Galaxy Evolution and environment
\label{subsec:evol_vs_env}}

The most massive galaxies known today are giant ellipticals that reside
in the most dense galaxy environments \citep[e.g.,][]{Dre:80}, which are
also known to have formed their stars rapidly at an early stage in cosmic
history. Recent studies of the evolution in the global space density
of galaxy SFRs find that the majority of the observed SF occurs in the
highest mass galaxies at high redshift, moving to lower mass galaxies
at lower redshifts \citep{Pan:04,Jun:05,Pan:06,Sey:07}. This is referred
to as ``downsizing'' \citep{Cow:96}. Together with the trend for massive
systems to occur predominantly in dense environments, this results in
a scenario where the SFRs in dense environments at low redshift are
lower than in less dense regions, consistent with the observations.
The implication is that at sufficiently high redshifts,
($z\gtrsim 2$) where the SFR is dominated by the most massive
galaxies, the SFR-density relation should invert, with the most
dense environments hosting elevated SFRs compared to the field.

This trend is beginning to be observed too. \citet{Pog:06} show that
at $0.4<z<0.8$ the suppression of SFR in dense environments is weaker than
seen locally. \citet{Ilb:06} and \citet{Elb:07} show environmental dependent evolution in
the galaxy luminosity function from $0.25<z<1.5$, suggesting an
increase in the density of faint red galaxies in overdense regions as
cosmic time increases. More confirmation is clearly needed, of course,
but these examples are consistent with the scenario that galaxies in dense
environments form stars rapidly at early times, quickly building up mass
and becoming quiescent, while galaxies in less dense environments form
stars at a more sedate pace but over longer timescales. 

This ``environmentally governed evolution" scenario makes a distinct prediction
compared to the ``infall and quench" models. The latter suggests that galaxies
in dense environments should show an SFR distribution that is progressively
suppressed from the outside in, as the outer regions are those which will be
affected first by their rapidly changing environment. The former,
on the contrary, suggests that the suppression should either happen
uniformly as a galaxy ages, or that the inner regions should be
suppressed first, since nuclear SF seems to occur more
rapidly than disk SF given the elevated gas densities present.
Thus by studying the spatial distribution of SFR in star-forming galaxies as
a function of environment we should be able to distinguish clearly between
these two scenarios.

\subsection{Radial Variation in Galaxy SFRs
\label{subsec:radial_var}}

Previous studies on the radial dependence of SF have focused on individual
galaxies. \citet{Per:06} studied the recent SF in the
early-type galaxy M81 using imaging observations from the far-ultraviolet
(UV) to the far-infrared (IR). The data was then compared to models of
stellar, gas and dust emission, with results from different sub-galactic
regions, including individual HII regions (around 0.1 kpc). They were able to
confirm the existence of a diffuse dust emission not directly linked
to the SF. Using the H$\alpha$ emission that probes the unobscured SF,
and the IR luminosity (especially the $24\,\mu$m emission) that probes
the obscured SF in the galaxy, they found a decrease in the ratio of
obscured SF to total SF with radius. This fraction varies from an
obscured SF of 60\% in the inner regions of the galaxy to 30\% in the
outer regions. \citet{Joh:05} used pixel-based spectral energy distribution
(SED) fitting to a merging system hosting a compact steep spectrum
radio source, in order to explore the connection between the nuclear radio
emission and the distribution of star formation. \citet{Kas:03} used
pixel-based colors to explore stellar populations and obscuration in the
Antennae, and \cite{deG:03} used the same technique to explore the Mice
and the Tadpole interacting galaxy systems.
\citet{Boi:06} showed, using GALEX and Spitzer data,
that for disk galaxies the attenuation varies radially, being highest
in the nuclear regions, and is correlated with metallicity. They also
found that the Schmidt law connecting the SF and gas surface densities
continues beyond the traditional ``threshold" radius. \citet{Lan:07} used a study of pixel Color Magnitude Diagrams for a sample of 69 nearby galaxies to study stellar populations and structure of galaxies. They found that these Color Magnitude Diagrams of each galaxy type have distinct trends. In addition, they performed a pixel-by-pixel analysis to show that there is a steady progression in average pixel color along the Hubble sequence. Finally they compared pixel colors to the Bruzual and Charlot stellar population models and used these to map the stellar mass distribution and M/L ratio in galaxies. 

The current analysis has the advantage of a very large sample, accessible
as a result of the SDSS, which allows a study of the environmental
dependence of spatially distributed star formation. Also, since the inferred
SFRs come from fitting SEDs, derived from stellar population synthesis models,
to broadband photometry, there is no aperture effect such as that affecting
fiber-based spectroscopy.  By fitting SEDs to individual pixels within
resolved galaxy images, we can spatially resolve the star formation
and explore how this distribution varies for galaxies as a function of
local galaxy density.

\subsection{Using photometric redshifts to study (spatially-resolved)
properties of stellar populations in galaxies
\label{subsec:photoz}}

Photometric redshifts have become an efficient way of measuring
redshifts of galaxies, particularly when the galaxies are too faint
for spectroscopic studies. This typically involves fitting a library of
SEDs to the observed colors of galaxies, where the SED templates are
functions of galaxy type. We can, however, shift the focus from the inferred
photometric redshifts themselves, to what the best fitting SEDs can
reveal about the properties of the galaxy sample. This highlights instead
the reliability and appropriateness of the template spectra used in the
fitting. More specifically, as shown by \citet{Abr:98}, the {\em spatially
resolved\/} colors of galaxies can be used to understand the
relative ages of bulge and disks and the formation histories of galaxies.
This can be done by applying the photometric redshift technique to
individual pixels of resolved galaxy images, having measured fluxes and
flux errors. Given a library of SED templates constructed from population
synthesis codes, for which the underlying physical parameters are defined,
the spatial distribution of those parameters can then be established.
\citet{Con:03} applied such SED fitting, a technique referred to as
``pixel-z," to individual pixels in $\approx 150$ galaxies with measured
spectroscopic redshifts, and $\approx 1500$ galaxies with measured photometric
redshifts, in the Hubble Deep Field-North (HDF-N). The aim was to
decompose the internal photometric structure of galaxies into intrinsic
properties of the stellar populations like stellar ages and star formation
rates. They used the pixels to calculate the comoving density of star
formation and metallicity enrichment, as a function of redshift. With
the sample of galaxies available, they were able to directly assess the
drivers behind the current understanding of the global star formation history.

Here we apply the pixel-z technique to 44,964 SDSS galaxies, to study
the dependence on environment of the total galaxy SFR and the spatial
distribution of star formation within those galaxies. In \S\,\ref{sec:data},
we describe our SDSS sample. We describe the pixel-z method and
its application to the SDSS in \S\,\ref{sec:method}, together with a description of the
SEDs and the range of physical properties spanned.
\S\,\ref{sec:implementation} demonstrates features of the pixel-z implementation.
\S\,\ref{sec:results} presents our results, including the relation
between total SFR and local density of galaxies, along with the spatial
distribution of SFR as a function of this density. Our conclusions are described in \S\,\ref{sec:conclusions}.
We assume throughout that $\Omega_{\Lambda}=0.7$, $\Omega_{\rm M}=0.3$ , and $H_{0}=75\,{\rm km\,s^{-1}\,Mpc^{-1}}$.

\section{Data
\label{sec:data}}

\subsection{The Fourth Data Release of the Sloan Digital Sky Survey
\label{subsec:dr4}}

The Sloan Digital Sky Survey (SDSS) is an imaging and spectroscopic survey
of the sky using a dedicated 2.5m telescope \citep{Gun:06} at Apache
Point Observatory in New Mexico. It aims to map a quarter of the sky,
spanning the Northern Galactic Cap \citep{Yor:00}. Imaging is done
in drift-scan mode using a 142 mega-pixel camera \citep{Gun:98},
which gathers data in five passbands, $u,g,r,i,z$, spanning wavelengths
$3000\lesssim \lambda \lesssim 10,000\,$\AA. The photometric system and
calibration are described in \citet{Fuk:96,Hog:01,Smi:02,Ive:04,Tuc:06}.
The astrometric calibration is described by \citet{Pie:03}
and the data pipelines in \citet{Lup:01}. The
Fourth Data Release (DR4) includes five-band photometric data for 180
million objects selected over 6670 square degrees, and 673,280 spectra
of galaxies, quasars, and stars selected from 4738 square degrees of
that imaging data using the standard SDSS target selection algorithms.
Objects are selected from the imaging data for spectroscopy using a
variety of algorithms. These include a complete sample of galaxies with
reddening-corrected \citep{Sch:98} \citet{Pet:76} magnitudes brigher 
than $r=17.7$ \citep{Str:02}.

\subsection{Sample Selection
\label{subsec:sample}}

We use galaxies from the Main Galaxy Sample and have a redshift
confidence of at least 0.7. We ignore objects with saturated pixels. We
then exclude galaxies whose spectra has the Z\_WARNING\_NO\_BLUE (no blue
side of the spectrum) and  Z\_WARNING\_NO\_RED (no red side of the spectrum)
flags \citep{Sto:02}. We then define a volume-limited
sample through the further restrictions $0.019<z<0.125$ and $-23.5<M_r<-20.5$.
This gives a sample of 44,964 objects. For each object we obtain
the atlas image (from the SDSS Data Archive Server), which comprises the
pixels detected as part of each object in all filters. We use these images
in the pixel-z analysis of the final volume-limited sample.

\subsection{Galaxy Environment
\label{subsec:environment}}

We characterize the local density around each galaxy using a $5\,$Mpc h$^{-1}$
sphere centered on  the galaxy in question.  Each galaxy within that
sphere is weighted according to the local completeness as calculated
by \citet{Bla:05} to account for spectroscopic fiber collisions.
Likewise, we scale the volume of the sphere according to the fraction
of the projected sphere contained within the survey area.  Since we
are using a volume limited sample, we do not need to correct for
redshift distribution variations.

\begin{figure}
\epsscale{0.75}
\plotone{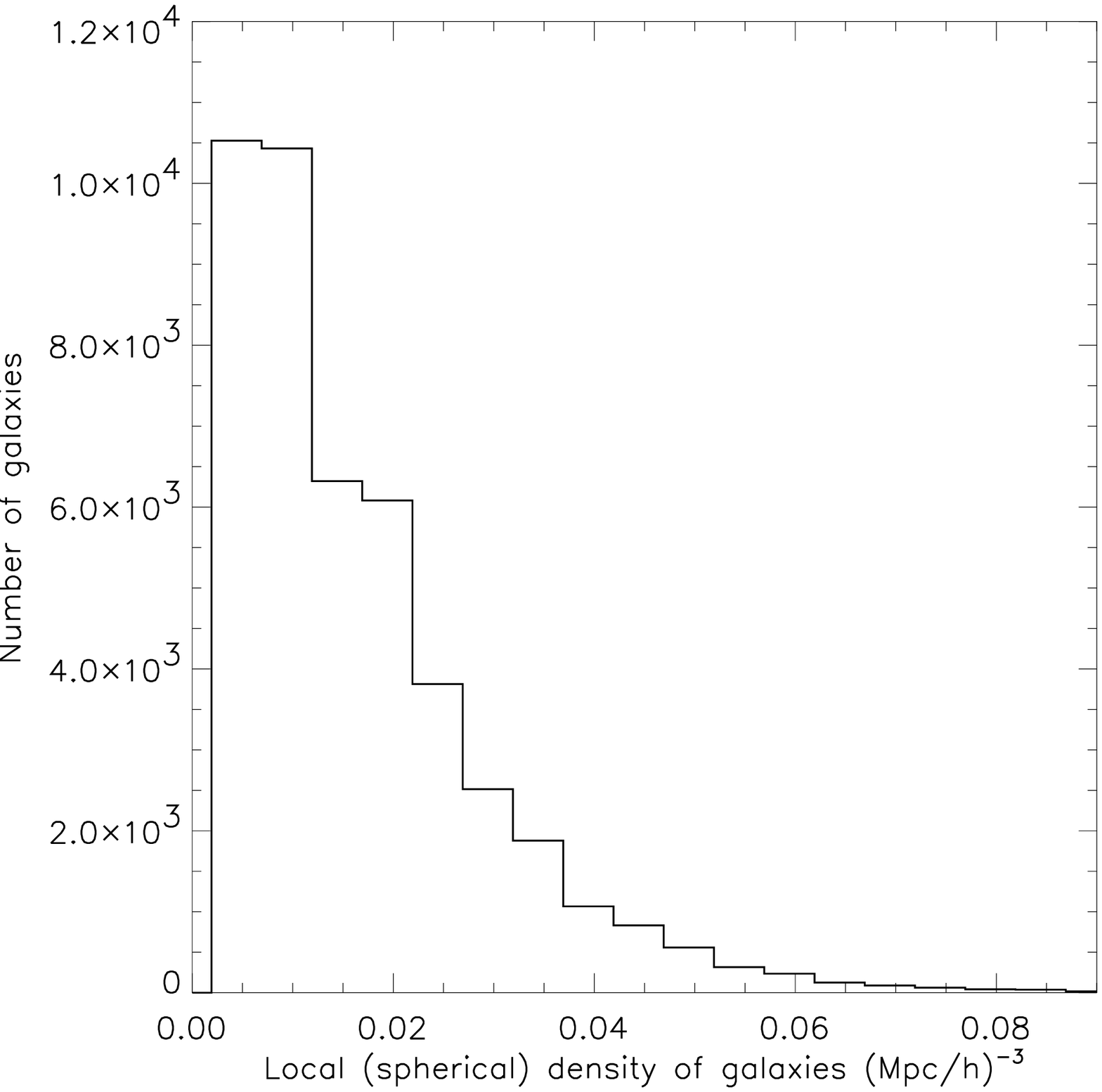}
\caption{Density distribution of galaxies in the sample.
\label{fig:densities}}
\end{figure}

There are a few possible caveats associated with this method.
Naturally, working in redshift space versus real space opens us up to
the possibility of galaxies scattering out of our volume in high
density regions where the peculiar velocities would be high and
vice-versa in lower density regions.  In addition, our calculation of
the sphere volume is compromised somewhat by treating masked regions
near the center of the projected sphere identically to those on the
edges, despite the fact that the former would remove a larger volume
from the sphere than the latter.  Of these, the redshift distortions
are more problematic; our sky coverage is sufficiently uniform that
tests using an exact spherical volume were not significantly different
from those using our method.  However, given that redshift distortions
are at least equally problematic for any density estimates working in
redshift space \citep{Dre:80,Gom:03}, we feel that the
more physically-based aspect of our method makes it the preferred
approach. Figure~\ref{fig:densities} shows the distribution of local galaxy densities in our sample.

\section{The method
\label{sec:method}}

\subsection{pixel-z 
\label{subsec:pixelzmethod}}

By making use of strong spectral features such as the $4000$\,\AA\
break, the Balmer break and the Lyman decrement, the standard
photometric redshift techniques can be used to quickly provide an estimate
of a galaxy's redshift. Each SED is systematically redshifted,
convolved with the photometric filter response functions, and compared
with the observed fluxes through each filter. In the pixel-z method, however, we assume a redshift and fit for the SED type. In fact, for the SDSS Main Galaxy Sample, all galaxies have measured spectroscopic redshifts, so $z$ can therefore be fixed in the fitting function. The best-fitting template is then established for each pixel at that particular redshift.  The fitting function has the form:
\begin{equation}
\label{eq:lkhood}
\chi^2(T) = \sum_{i=1}^{N_{\rm f}}
\frac{[F_{{\rm obs},i} - b_j \times F_{i,j}(T)]^2}{\sigma^2_i}
\end{equation} 
$F_{{\rm obs},i}$ is the flux through the $i$th filter, $b_j$
is a scaling factor, $F_{i,j}$ is the flux through the $i$th filter
of the $j$th spectral energy distribution template (calculated at redshift
$z$) and $\sigma_i$ is the uncertainty in the observed flux. The sum is
carried out over all available filters $N_{\rm f}$.
The resulting $\chi^2$ is minimized as a function of template $T$ providing an estimate of its spectral type (together with the variance on this measure). 
Minimizing $\chi^2$ in equation~\ref{eq:lkhood} with respect to $b_j$ gives
\begin{equation}
\label{eq:scale}
b_j(T) = \frac{\sum_{i=1}^{N_{\rm f}}\frac{F_{{\rm obs},i}F_{i,j}(T)}{\sigma_i^2}}{\sum_{i=1}^{N_{\rm f}}\frac{F_{i,j}(T)^2}{\sigma_i^2}},
\end{equation}
which determines the normalization of the SFR obtained from the
best fitting SED template.

Rather than applying this technique to the integrated fluxes of galaxies,
we instead apply it to the fluxes of pixels within resolved galaxy images.
The individual pixels typically have larger photometric uncertainties than
integrated fluxes measured in apertures, and careful account needs to be made
of the uncertainties and error-propagation. The optimum solution would
be to keep the spatial information present in the resolved image together
with the improved signal-to-noise ratio offered by combining pixels.
A step in this direction is indicated by \citet{Con:03}
(T. Budav{\'a}ri, 2003, private communication), where spatially connected
pixels of similar colors are joined into {\em superpixels} in
order to improve on the statistical errors without mixing the different
galaxy components, such as a red bulge or bluer SF regions in spiral arms.
Incorporating this technique into the current pixel-z implementation is
beyond the scope of this investigation, but holds promise for future work.

By careful choice of the SED template library, the pixel-z technique enables
a decomposition of the internal photometric structure of galaxies
into basic constituents such as the age of the stellar population,
their metallicities and their dust content. It enables, under simplifying
assumptions, the determination of the SFR for individual pixels
inside a galaxy, and the contribution of each pixel to the SFR
of either the whole galaxy or a projected radial shell of that galaxy.

We have thus shifted the attention of the technique from the photometric
redshift itself to the SED templates themselves. In fact, the SED of
a galaxy should reflect the distribution of stellar masses, ages and
metallicities and hence provide clues to the past history of star
formation. By fitting SEDs to individual pixels in a galaxy, we can
recover the morphological characteristics of the galaxy and separate the
individual contributions of age, metallicity, dust and star formation
history.

To verify our implementation of pixel-z and for comparison with the
initial application to the HDFN by \citet{Con:03}, we test our method
first on HDFN galaxies before moving to our SDSS sample. As in the analysis of \citet{Con:03}, we are able to connect features in the parameter maps (of age, SFR, obscuration, metallicity) of galaxies to individual morphological features such as knots of star formation that appear in the original images. The HDFN images have accompanying rms maps that quantify the uncertainty due to
noise in the background. These maps were used to determine the detection
significance of an object or a pixel. No such maps are available for
the SDSS galaxies. The photometric calibration of the SDSS imaging data (Atlas images) is done using the {\em asinh} magnitude system according to \citet{Lup:99}. The photometry is also corrected for foreground Galactic
extinction using the extinction values obtained from the dust maps of \citet{Sch:98}.

\subsection{SED Templates
\label{subsec:templates}}

We use a large number of SEDs generated by the \citet{BC:03} stellar population synthesis models. The main input parameters are the form of the SFR, the stellar
initial mass function (we assume a Salpeter function with
$M_l=0.1\,M_{\odot}$ and $M_u=100\,M_{\odot}$), and the rate of metal
enrichment. The SEDs are normalized to a total mass of $1\,M_{\odot}$.
The input parameters for our SED templates are chosen to maximize
our ability to solve for the above quantities. The SEDs have the
following properties:

\begin{enumerate}
\item We allow the underlying stellar population within each pixel
to vary over a wide age range. The age in the synthesis model is defined as the
time since the most recent burst of star formation. We sample extremely young
($0.001, 0.01, 0.1, 0.5$ Gyr), to middle age ($1,3,5$
Gyr), to old and very old ($9, 12, 15$ Gyr), for a total of ten ages.
\item We assume that the fluxes of individual pixels can be
modeled using an exponentially declining SFR with an $e$-folding
timescale $\tau$, i.e.\ $\Psi(t) = \Psi_0 e^{(-t/\tau)}$. This
parameterization is convenient for its simplicity in describing the SFR
of an instantaneous burst when $\tau\to 0$ and a constant SFR
when $\tau\to\infty$. The $e$-folding times we use for $\tau$ range
from $0.1$\,Gyr for a short burst, to $1,3,5,9$ and $12$\,Gyr for
subsequently longer bursts. It is worth noting at this point that an
exponential SFR for individual pixels does not inevitably lead to an
exponentially decaying SFR for the galaxy as a whole, other than in the
special case in which every pixel in the galaxy is coeval and all have
the same SFR.
\item Since pixels with any SF history can be expanded in a series
of instantaneous bursts, each having fixed metallicity,
the spectral evolution of individual pixels (or whole galaxies) can be
investigated without prior knowledge of chemical evolution. We
assume the SEDs to be characterized by six possible metallicities
spanning $\frac{1}{50}$ to $2.5$ solar.
\item The general spectral characteristics of the SEDs of galaxies will
be modified by the presence of dust. We parameterize dust obscuration
in terms of the relative optical extinction in the rest frame $E(B-V)$
using the reddening curve $k(\lambda) = A(\lambda)/E(B-V)$, for
star-forming systems formulated by \citet{Cal:00}. For each of the SEDs
we allow for six independent values of extinction ranging from no
extinction to $E(B-V)=0.9$ magnitudes of extinction.
\end{enumerate}

\section{Pixel-z Implementation
\label{sec:implementation}}

We fit 2160 SED templates to all pixels in each of the SDSS galaxies in
our sample, keeping the redshift of all pixels fixed to the spectroscopic
redshift of their host galaxy. As discussed earlier, this effectively
removes one degree of freedom in the fit and returns the properties
of each of the pixels in terms of their best-fitting template, i.e.\
the age, SFR $e$-folding time, dust obscuration and metallicity.
The normalization of the SFR in each pixel is based on the scale factor $b_j$ which gives the SFR in units of $M_{\odot} \,yr^{-1}$.  
The total SFR for each galaxy is
calculated from the sum of all pixels in the galaxy. We construct both
unweighted and weighted sums to give a total star formation rate for the
galaxy, where the weights correspond to the reciprocal of the square of the fractional error on the
SFR in each pixel. The weighting is done to avoid giving undue significance to poorly constrained pixels, such as pixels dominated by the sky background, so that these pixels do not bias our measurements. The unweighted results turn out to be qualitatively the same as the weighted ones, but with some quantitative differences.

In calculating the total SFR we remove the simplifying
assumption used by \citet{Con:03} that all the pixels are coeval, i.e that
they share a common age over the whole galaxy. The assumption of common age
simplifies the interpretation by removing a degree of freedom, and enabling
a clearer vision of the interplay between the remaining fitted parameters.
That approach was useful in the analysis of \citet{Con:03} in comparing
maps of SFR, dust obscuration and metallicity for each galaxy with the
underlying morphology. This assumption, however, is not consistent with
the galaxy formation scenarios we are testing, nor is it required for our
current analysis, and so we dispense with it.

\subsection{Galaxy Maps}
\label{subsec:maps}

The technique allows us to probe not only the underlying morphological
details of the galaxy, but also the relation between a galaxy's morphology
and its physical constituents.

\begin{figure}
\epsscale{1.0}
\plotone{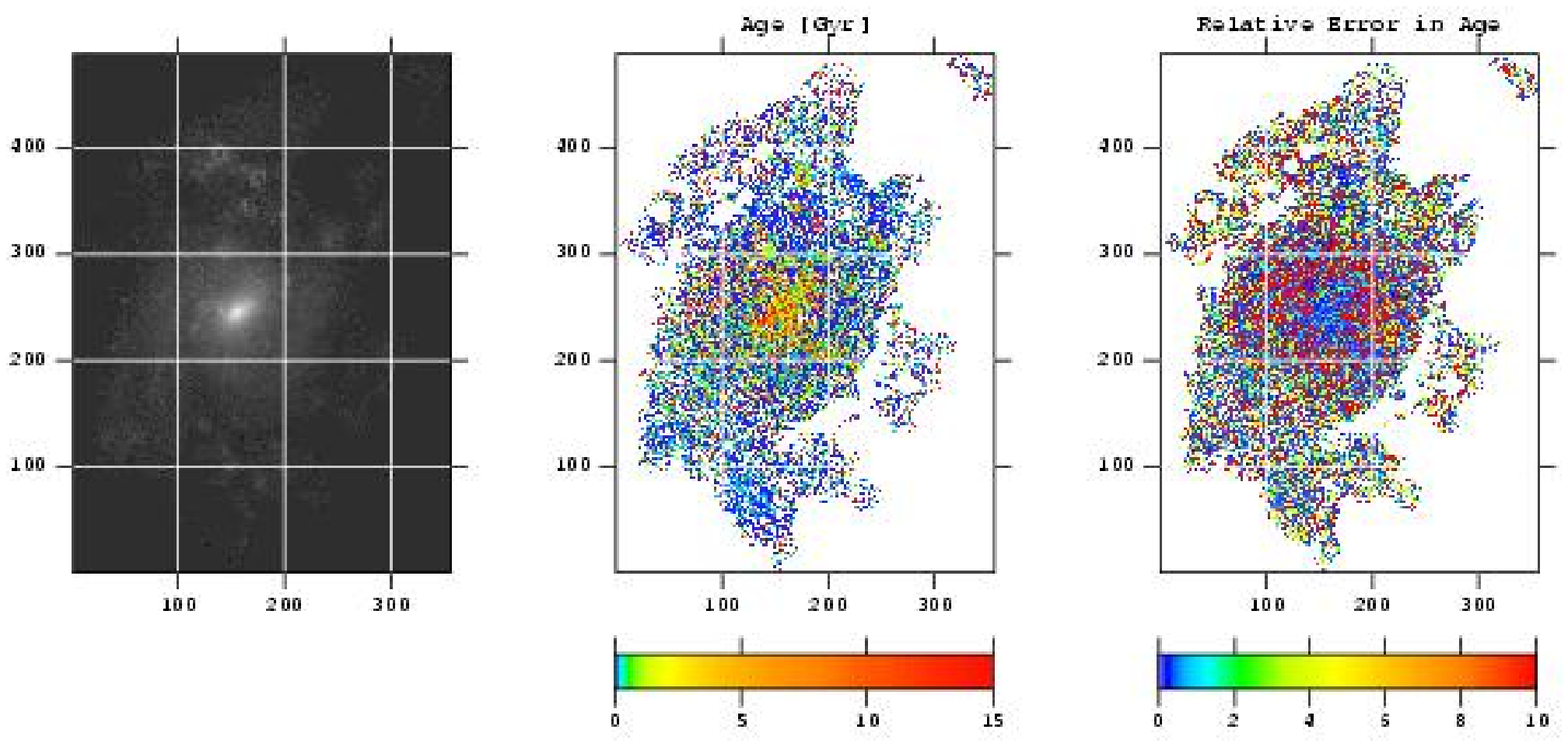}
\plotone{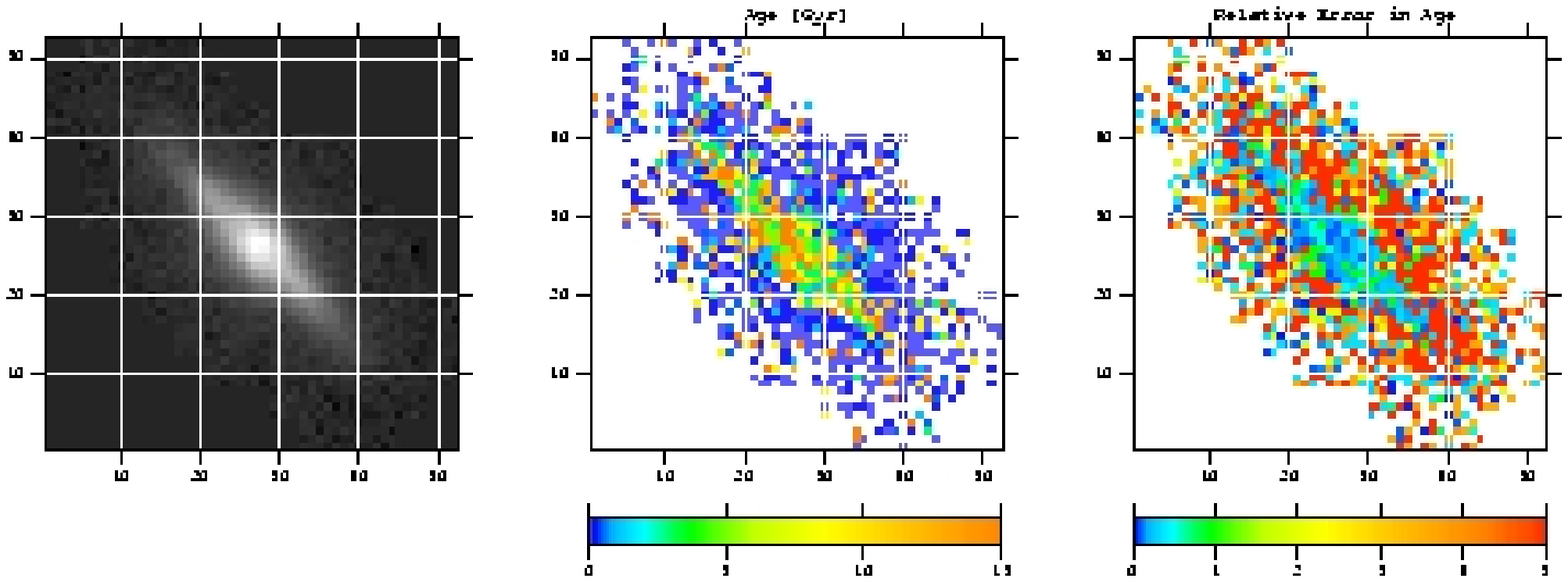}
\caption{
Top panel: Result of fitting 2160 SED templates to NGC 450 (SDSS J011530.44-005139.5) in the SDSS. The top left map shows the galaxy in the SDSS r' filter. The middle map displays the breakdown of the best-fitting template in each pixel according to the values of stellar population age in Gyr. Blue regions have the lowest age (0.001-0.1 Gyr), then green (0.1-0.6), then yellow (0.6-2 Gyr), then orange (2-7 Gyr), then red (7-15 Gyr). The third map shows the corresponding relative errors in the age for the pixels in this galaxy. As can be seen, the relative errors increase when going from the bulge region to the spiral arms.
The map is obtained by computing the marginalized likelihood for each pixel in the galaxy as described in section 4. Bottom panel: Another galaxy (SDSS J155919.97+061729.8). The left image shows the image in the r' filter. The middle map is the age map and the right image is the corresponding relative error map on the age.
The nuclear region, with high flux pixels, shows a lower relative error on the age but the disk has a higher relative error. A substantial number of sky pixels are captured on the outskirts of this galaxy. These have the highest relative errors.
\label{fig:SDSSGalaxies}}
\end{figure}

The first two panels of Figure~\ref{fig:SDSSGalaxies} show the results of the decomposition for two galaxies. The image on the left of the top panel shows a
SDSS spiral galaxy, NGC~450 (SDSS J011530.44-005139.5) and the middle image shows the distribution of best fitting stellar population ages in Gyr throughout this galaxy. We see an older population in the nucleus of the galaxy  (ranging from 5-12 Gyr). In the outskirts there is a younger population, whose ages range from 0.001 to 0.2 Gyr. 
The distribution of this parameter also traces out some of the spiral arm
structure. Around the inner spiral arms, some of the ``knots'' of star
formation seen in the original $r'$ band image can be detected in the
age image too as yellowish-green regions (with ages between 0.1 and 2 Gyr).
The bottom panel shows the results for an edge-on disk galaxy. We detect an older population in the nuclear region (around 12 Gyr) and a mix of intermediate (2-5 Gyr) and old populations (5-12 Gyr) elsewhere in the disk. The stellar populations in the outskirts have a comparitively younger age.

\begin{figure}
\epsscale{1.0}
\plotone{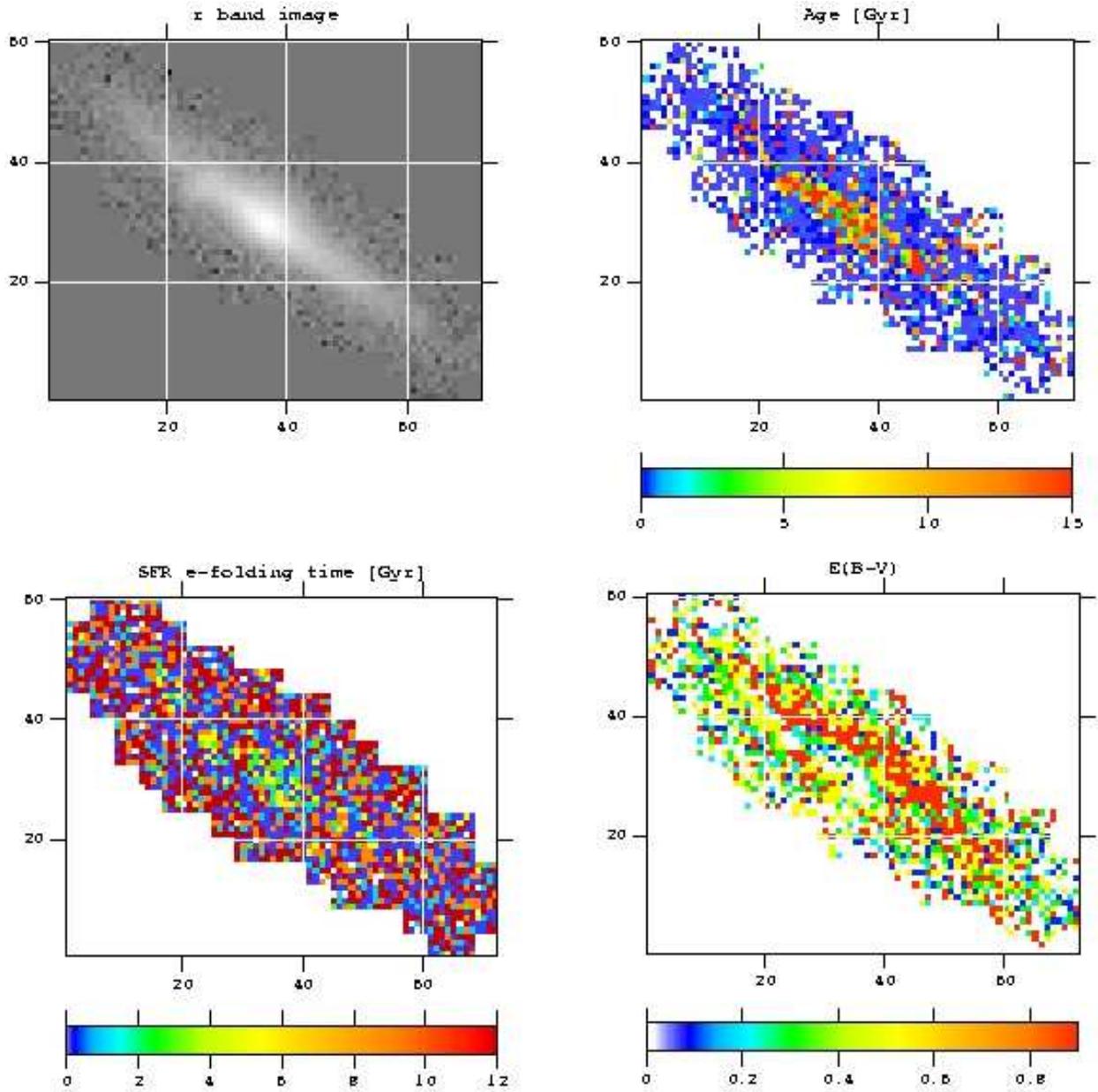}
\caption{Result of fitting 2160 SED templates to each pixel in the SDSS galaxy  SDSS J075642.69+364430.0. The top left map shows the galaxy in the SDSS r' filter. The other three maps display the breakdown of the best-fitting template in each pixel according to the values of the age of stellar population, star formation rate e-folding time in Gyr and color excess in magnitudes.
\label{fig:sdss2}}
\end{figure}

Figure~\ref{fig:sdss2} is a result of the template decomposition for another SDSS galaxy (SDSS J075642.69+364430.0).
There is evidence for a bulge in this disk galaxy - the stellar
populations are older - as much as 10 to 15 Gyr, whereas those in the disk are much younger, typically less than 1 Gyr. The population in the core also has lower $e$-folding time ($4<\tau<5$ Gyr) than in the outer disk. Together these results point to a lower current SFR in the central (bulge) population, compared to the disk population.
The nucleus itself has a relatively low obscuration with $E(B-V)\approx 0.0$, while the stellar populations in the outer parts of the disk typically show a higher obscuration ($E(B-V)\approx 0.9$). The central part of the galaxy therefore shows an older population of stars, a faster timescale for SFR decline and a lower level of obscuration than the stellar populations in the outskirts of the disk. As we go further out we expect to be dominated by the sky pixels, which are artificially best fit by our SED templates corresponding to a younger stellar population, a longer timescale for SFR decline and a higher obscuration. Again, this artifact is identified through the large errors in the fitting, an important aspect of the pixel-z analysis, discussed in detail next.

\subsection{Error Maps}
\label{subsec:errmaps}

The pixel-z method includes calculation of the intrinsic error arising from
the SED fitting for each property characterizing the best fitting SED, as detailed by \citet{Con:03}. We provide a brief summary of the process here. Each pixel has a 4-dimensional likelihood function that results from fitting
each of the 2160 templates to the five band fluxes in that pixel. In
order to calculate the uncertainty associated with each of the four axes
of variability (age, SFR $e$-folding time, obscuration and metallicity),
we marginalize the likelihood over the three remaining parameters, essentially
collapsing the four dimensional function along each of its axes.
These likelihoods are sampled at the allowed values of each parameter.
For example, the age likelihood is sampled at 10 different points
corresponding to the 10 different ages.

The errors are then the $1\,\sigma$ uncertainties (corresponding to
$\Delta \chi^2=1$) found by measuring the width of the $1\,\sigma$ line that
intersects the curve. 
Each pixel thus has uncertainties associated with each of the four parameters.
In the top rightmost panel of Figure~\ref{fig:SDSSGalaxies}, we display the relative error in the age of the stellar populations. The inner arms of the spiral galaxy have a higher relative error in the age than does the central bulge region. In the second galaxy in the lower rightmost panel, the central bulge region has a lower relative error in the age than does the disk region.
 
\begin{figure}
\epsscale{1.0}
\plotone{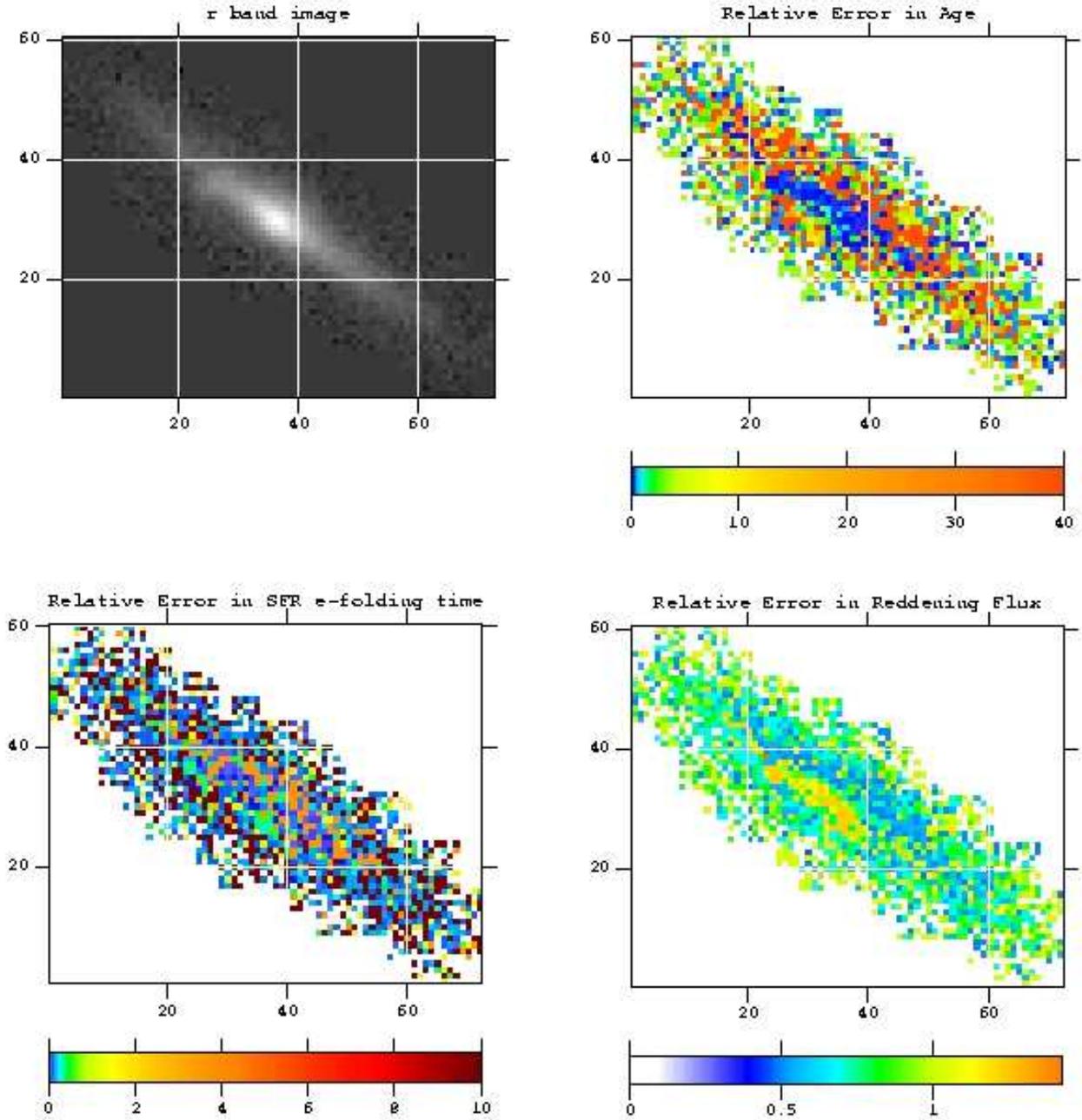}
\caption{Relative error maps for the galaxy shown in Figure \ref{fig:sdss2}. Original image (top left), relative error in age (top right), relative error in SFR $\tau$ map (bottome left) and relative error in reddening flux $10^{0.4(\Delta\,E(B-V) - E(B-V))}$ (bottom right). 
\label{fig:sdss2_err}}
\end{figure}

In Figure~\ref{fig:sdss2_err} the bulge region, which shows an older stellar population 
has a lower relative error in the age ($\frac{\Delta\,t}{t}\approx 0.1\,$) than the disk region which shows younger stellar populations with much higher relative uncertainties. These higher uncertainties correspond to lower flux pixels. The high values of these relative errors can be reduced somewhat by finer sampling of the age parameter near the low best-fitting age value. Now beyond the disk, the pixel-z technique artifically fits to the sky background, so that the seemingly lower values of the relative error on the age parameter in the outermost pixels are an artefact of the fitting and could also be a product of template degeneracy (discussed below). However, in either the disk or the sky, pixels with a relative error $>1$ will have a negligible contribution to the calculated SFR compared to those with very low relative errors. 

The bulge also shows a faster $e$-folding time for star formation, with a lower relative error ($0 < \frac{\Delta\,\tau}{\tau} < 0.4$) than in the disk ($\frac{\Delta\,\tau}{\tau} > 1.0$). The pixels in the outer regions within the galaxy generally show the highest relative errors associated with the age and $e$-folding time. This emphasises that calculation of total or local SFR has to be weighted by these fitting uncertainties. Similar trends are seen in other parameters though the bulge population in the obscuration map has a higher relative error in the obscured flux (with $10^{0.4(\Delta\,E(B-V) - E(B-V))}\approx 1.2\,$) than does the disk ($\approx 0.5\,$), although the relative error becomes large again in the furthest outskirts of the galaxy. 

In calculating either the total SFR of the galaxy, or the mean SFR in radial
shells, we weight the SFR in each pixel by its relative uncertainty as
calculated from the error maps for the age and $\tau$ parameter.
The weight of each pixel is proportional to the reciprocal of the
fractional error squared in the calculated SFR of each pixel.

\subsection{SED degeneracies}
\label{subsec:seds}

\begin{figure}
\epsscale{0.65} 
\plotone{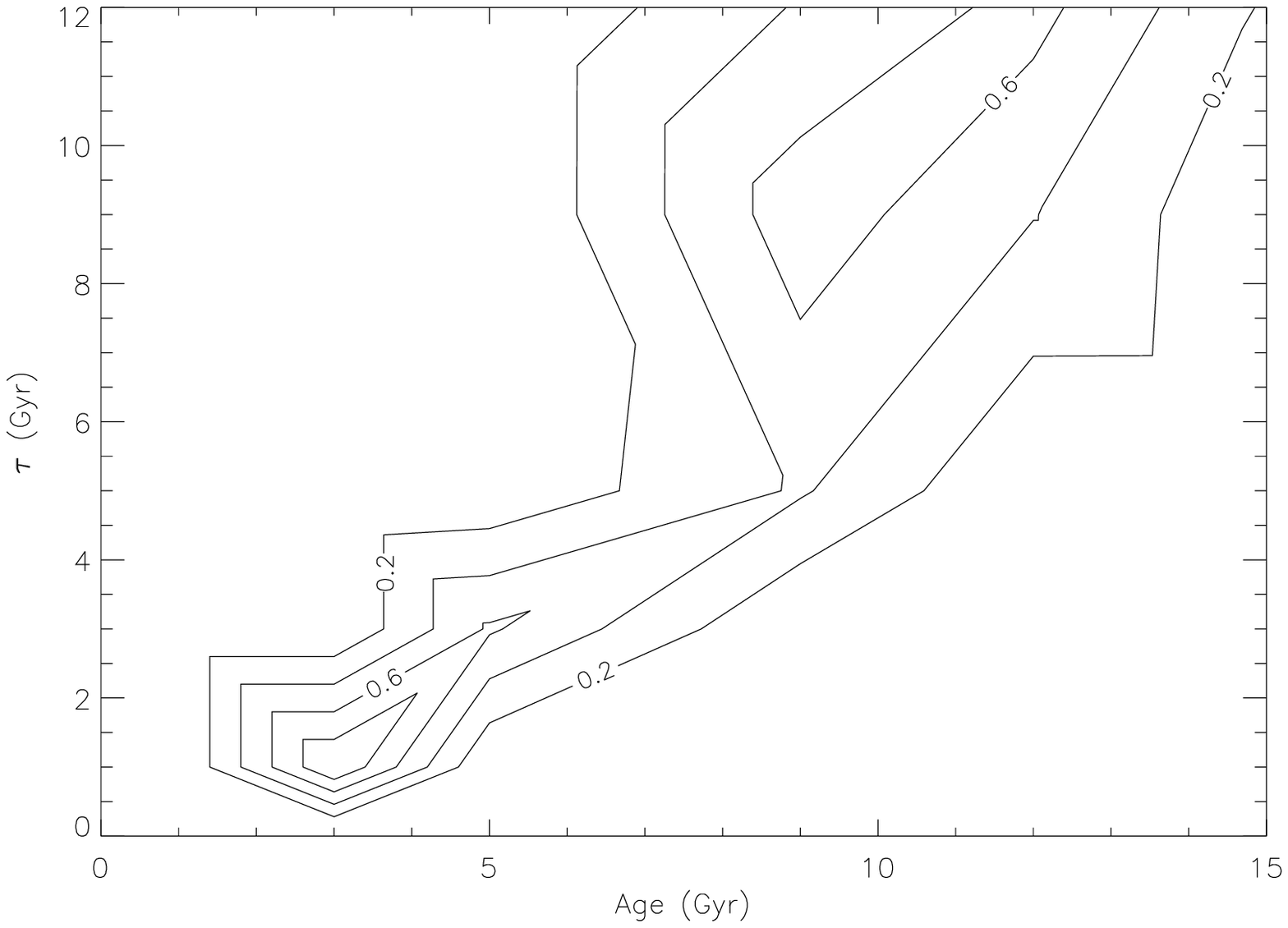}
\plotone{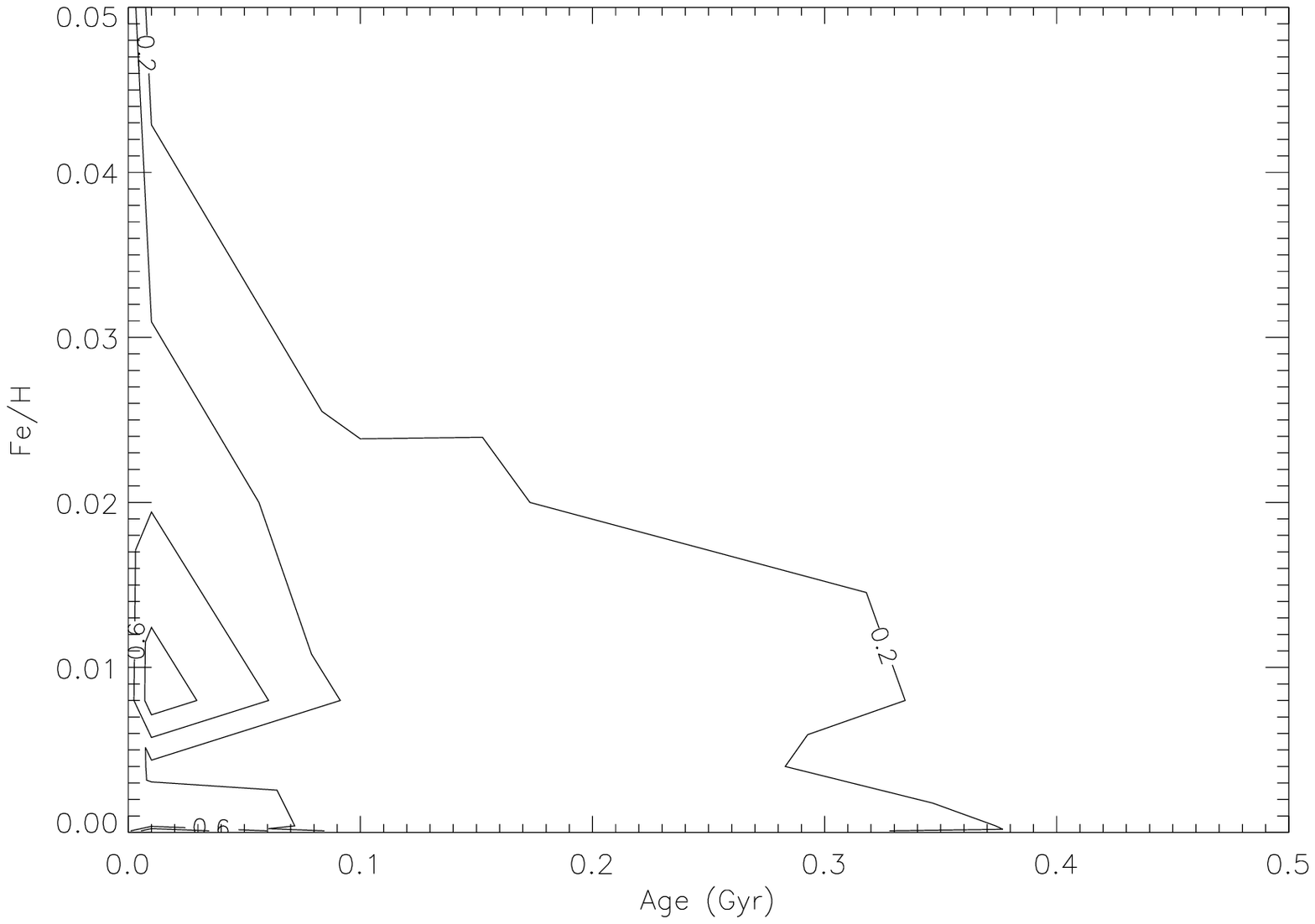}
\caption{Top panel: Age-$\tau$ likelihood contours for a pixel in the nucleus of a galaxy. Bottom panel: Age-metallicity likelihood contours for a pixel in the nucleus of another galaxy. Contours are obtained from marginalizing the 4 dimensional likelihood function onto these two dimensions. 
\label{fig:degeneracies}}
\end{figure}

Another uncertainty in the method is introduced by the degeneracy between
the different parameters. There is likely to be a correlation between the parameters
that determine the best-fit SED. We can investigate these correlations by
using the 2-dimensional marginalized likelihood function, i.e.\ collapsing
the 4-dimensional function onto the two axes of interest. Example
results from this analysis are shown in Figure~\ref{fig:degeneracies}. The
top panel shows the likelihood contours as a function of age and star
formation rate $e$-folding time for a single central pixel in an SDSS galaxy. This is typical of the degeneracies seen in the fitting of most pixels. The near-elliptical contours indicate that these two quantities are not
independent at least in the central region. The likelihood function is
double-peaked, with the global maximum corresponding to a 3\,Gyr old
population and a 1\,Gyr SFR $e$-folding time. As both these quantities
are used in determining the star formation rate, we could expect a
certain degeneracy in the calculated star formation rate. The bottom
panel shows the degeneracy between age and metallicity in a central pixel of another galaxy, with the
maximum corresponding to a 0.01\,Gyr old populations and a $Fe/H\approx 0.01$.
Although we have identified such degeneracies, we have not yet exhaustively
analysed their impact through all pixels in all galaxies in
the sample. The tests we have explored, though, suggest that their impact
will be minimal averaged over large samples of varied galaxy types.

\begin{figure}
\epsscale{0.75}
\plotone{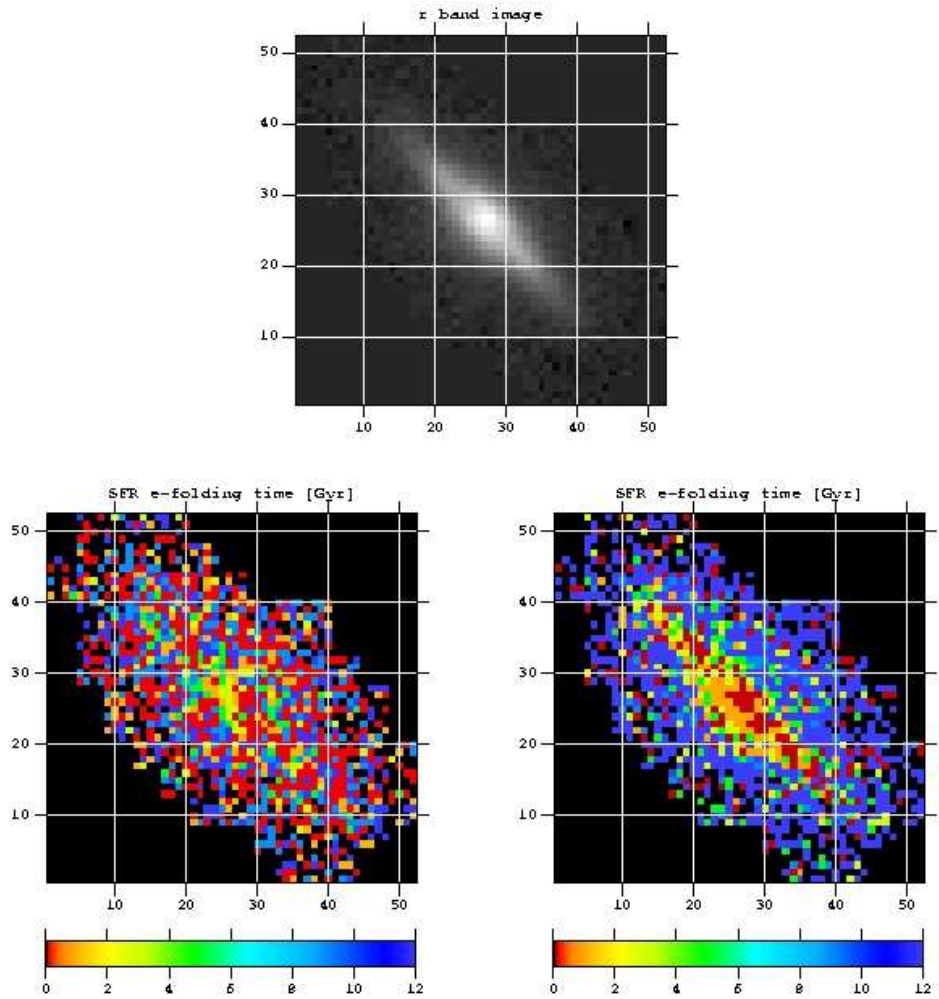}
\caption{Effect of age-$\tau$ correlations. Top panel: Image of galaxy in the r' band. Lower-left panel: $\tau$ map. Lower-right panel: $\tau$ map where all the pixels are constrained to have a common age. The reduced number of degrees of freedom enables a more resolved distribution of central bulge and disk in this $\tau$ map. 
\label{fig:agetaucorrelations}}
\end{figure}

One of the effects of this can be seen directly in the images. Figure~\ref{fig:agetaucorrelations} shows the effect of constraining all the pixels to have a common age. The lower-right image corresponds to the case where no constraint was used and a fit in the full parameter space was done, but the right image shows the case where the common age assumption was used. The reduced number of degrees of freedom enables a more resolved distribution of central bulge and disk in this $\tau$ map. 
Some of these degeneracies represent real physical relationships between the parameters, such as the well know age-metallicity degeneracy, while others reflect the choice of templates. Our selected templates cover the majority of parameter space, and since the fitting is done simultaneously over all parameters, the existence of degeneracies between any two parameters is mitigated. A more detailed study of the effect of these degeneracies is underway and will be incorporated in future work.

\section{SFR variation with local density
\label{sec:results}}

\subsection{Total galaxy SFR as a Function of Local Galaxy Density for All Galaxies}
\label{subsec:totsfr}

We investigate how the distribution of galaxy SFRs changes as a function
of the local (spherical) galaxy density. The total SFR in each galaxy is
calculated as a weighted sum of the SFR over all the pixels in the galaxy,
where the weights are inversely proportional to the square of the fractional
uncertainties in the SFR of each pixel. The local environment is quantified by counting the number of galaxies in a $5\,$Mpc\,h$^{-1}$ radius sphere centered on each galaxy.

\begin{figure}
\epsscale{0.65}
\plotone{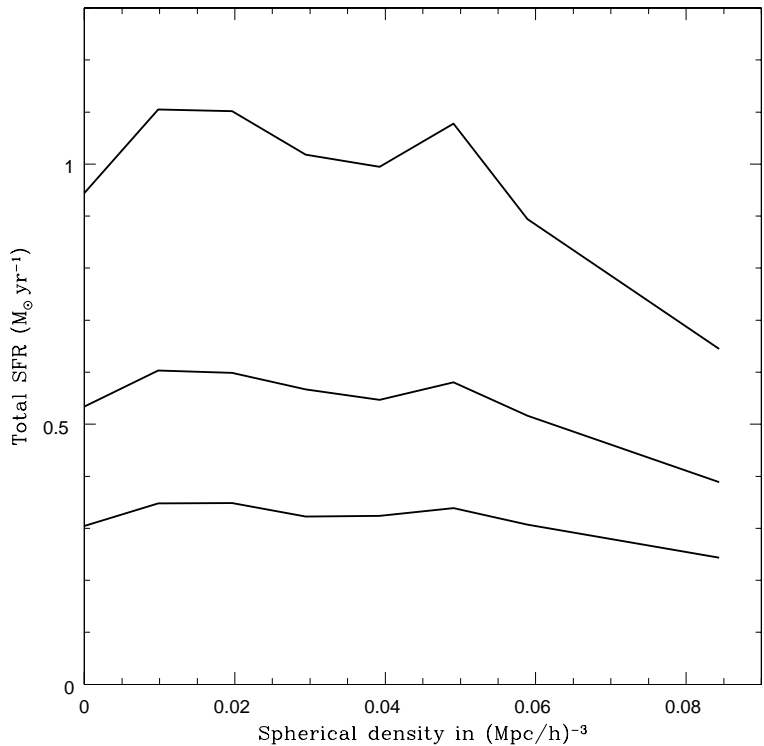}
\caption{Galaxy SFR in $M_{\odot} \, yr^{-1}$ as a function of density, where the SFR is calculated from a weighted sum over pixels in each galaxy. From top to bottom, the lines indicate the 75th, median and 25th percentiles of the SFR distribution respectively. The fluctuations at low densities correspond to the size of systematic uncertainties in the measurements. The overall SFR distribution which is essentially flat (given the size of these fluctuations) for low densities shifts to lower values at the higher densities. The decrease at higher densities is most noticeable in the most strongly star-forming galaxies, those in the 75th percentile. Finally, the range in the SFR distribution is largest at lower densities and decreases continuously with density. 
\label{fig:totalsfr}}
\end{figure}

Figure~\ref{fig:totalsfr} shows the variation of the SFR distribution
with local density, the three lines correponding to the 25th, median
and 75th percentiles of the SFR distribution. The fluctuations at low
densities are characteristic of the size of the systematic uncertainties in
these measurements. The SFR decreases with increasing density,
with the greatest effect in the highest density environments,
$>0.05\,$(Mpc/h)$^{-3}$ (densities that correspond to the outskirts of
rich clusters). 

These results, consistent with the measurements of \citet{Lew:02} and
\citet{Gom:03}, suggest that the total SFR of galaxies in the SDSS is
strongly correlated with local density. This is also in agreement with the
predictions of various hierarchical galaxy formation models whereby SF
in galaxies is suppressed as galaxies fall into more dense environments
such as clusters. The range of local densities in our sample allows us
track the total SFR in galaxies in a wide range of environments: from the
cores of rich clusters and groups into the field. As detailed by
\citet{Gom:03}, based on the SDSS Early Data Release and using H$\alpha$ EW as
a measure of SFR, Figure \ref{fig:totalsfr} illustrates the effect of
environment on SFR. The overall SFR distribution, essentially flat for
low densities, shows a decrease or suppression in regions of higher density. The effect is most noticeable
in the most strongly star-forming galaxies, i.e.\ those in the 75th percentile
of the SFR distribution. This means that the skewness of the distributions
decreases with increasing density. The ``break" density, beyond which the SFR
distribution falls rapidly to lower values, around $0.05\,$(Mpc/h)$^{-3}$,
occurs well into the regime of rich clusters, the extreme tail of the galaxy
density distribution. A comparison between the ``break" density seen here using our density estimator and that measured by \citet{Gom:03}
will be explored in future work.

\subsection{Effect of the Density-Morphology Relation: Total SFR-density Relation for Early and Late Type Galaxies}
\label{subsec:totsfr_types}

It is well established that in more dense environments the galaxy population becomes dominated by early-type galaxies. For example, \citet{Dre:80} through the study of 55 nearby galaxy clusters found that the fraction of elliptical galaxies increases and that of spiral galaxies decreases with increasing local galaxy density in all clusters. This indicates that the physical mechanisms that depend on the environment of each galaxy mainly affect the final configuration of stellar component. The density-morphology relation was also found in groups of galaxies. For example, \citet{Pos:84} using data from the CfA Redshift Survey, found that the density-morphology relation for groups is consistent with \citet{Dre:80}. The relation was also observed in X-ray selected poor groups \citep{Tra:01}. 

\begin{figure}
\centering
\includegraphics[width=.63\textwidth,height=0.46\textwidth]{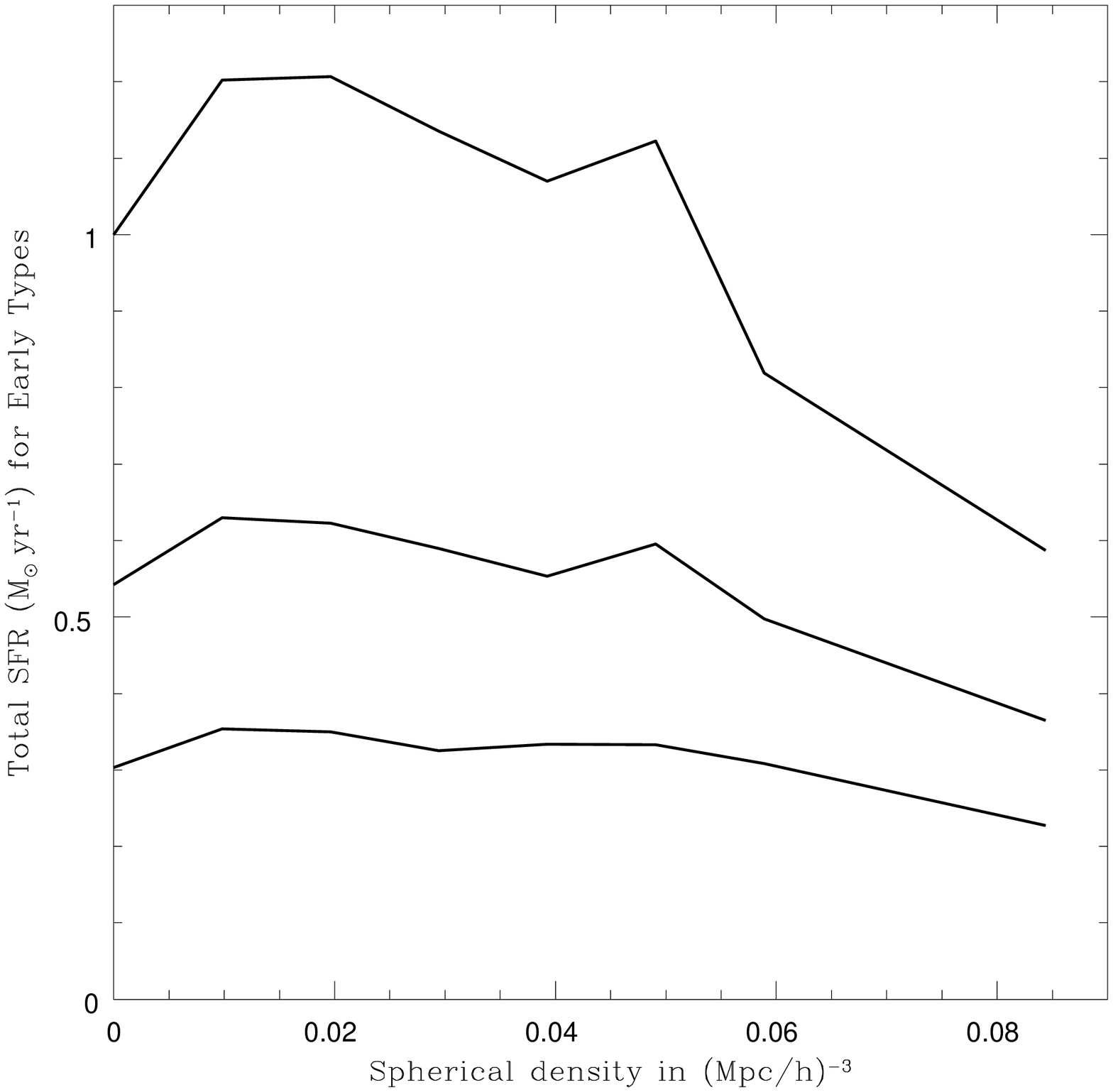}
\includegraphics[width=.63\textwidth,height=0.46\textwidth]{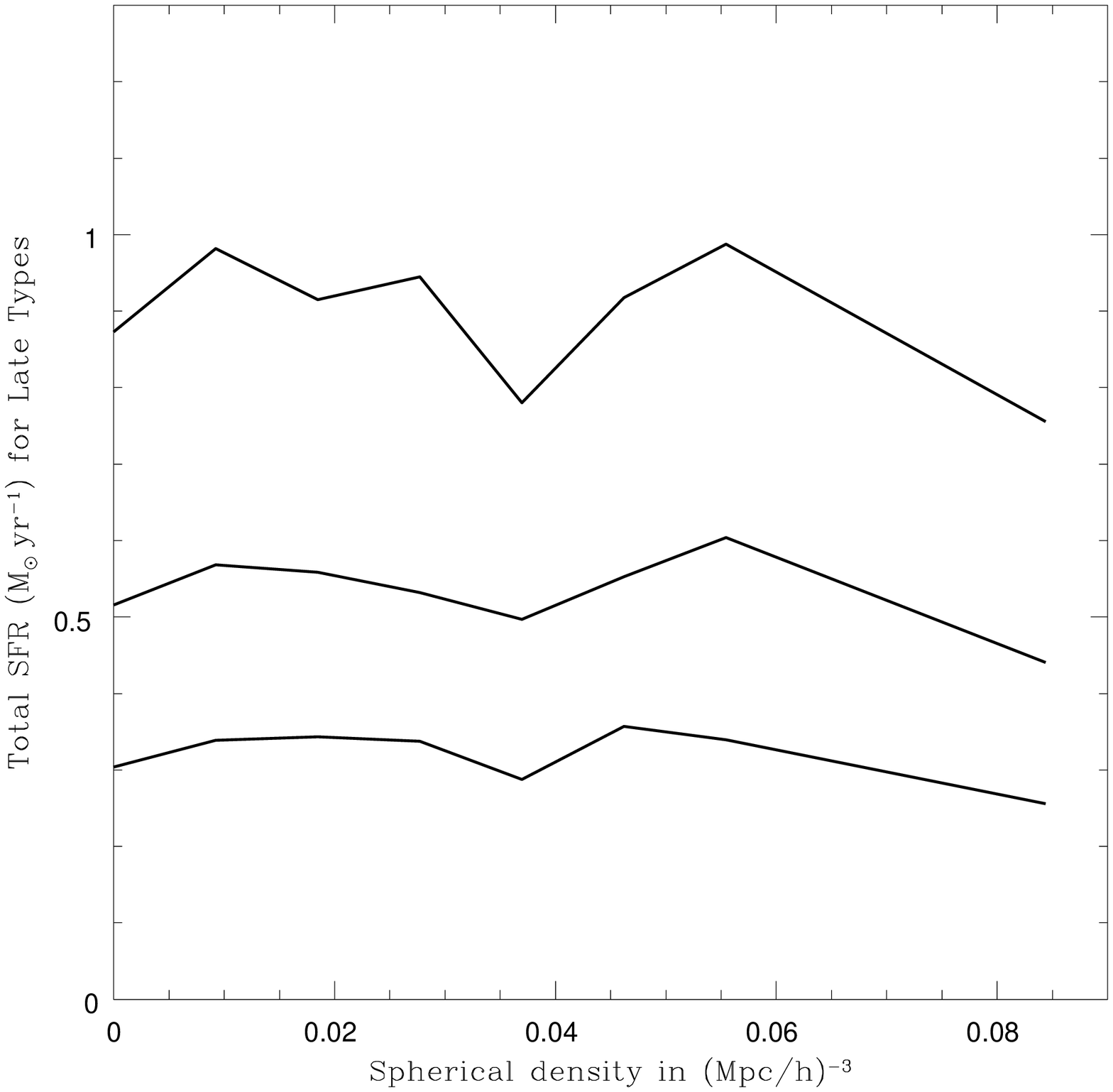}
\caption{Top panel: SFR-density relation for early-type galaxies (with $C \le 0.4\,$ as discussed in \S\,\ref{subsec:totsfr_types}). Bottom panel: SFR-density relation for late-types (with $C > 0.4\,$). In each panel, the total galaxy SFR is calculated from a weighted sum over pixels in each galaxy. From top to bottom, the lines indicate the 75th, median and 25th percentiles of the SFR distribution respectively. For early-types, the overall SFR distribution is relatively flat (compared to the size of systematic fluctuations) for low densities and then shifts to lower values beyond $0.05\,$(Mpc/h)$^{-3}$. The decrease at higher densities is most noticeable in the most strongly star-forming galaxies in the 75th percentile of the SFR distribution. For late-types, the overall SFR distribution is also relatively flat at low densities but then falls to lower values beyond $0.055\,$(Mpc/h)$^{-3}$. The range in the SFR distribution decreases with increasing galaxy density for early-types but is relatively unchanged for late-type galaxies.
\label{fig:totalsfr_types}}
\end{figure}

We seek to determine whether the SFR of galaxies of a given morphology are also affected by environment, i.e. whether the SFR-density relation holds regardless of mophology. We thus split our sample into two broad morphological bins based on the (inverse) concentration index $C$ i.e early-types with $C \le 0.4\,$ and late-types with $C > 0.4$. Consequently, there are 27,993 early-type galaxies and 16,971 late-type galaxies. Figure~\ref{fig:totalsfr_types} presents the distribution of SFR as a function of the local galaxy density for each of these morphological types. For early-types, the density-SFR relation is similar to the one observed for the full sample: the overall SFR distribution is relatively flat (compared to the size of systematic fluctuations) for low densities and then shifts to lower values beyond $0.05\,$(Mpc/h)$^{-3}$. Like the trend for the full sample, the decrease at higher densities is most noticeable in the most strongly star-forming galaxies in the 75th percentile of the SFR distribution, with a sharp decrease beyond $0.05\,$(Mpc/h)$^{-3}$. The scatter in the SFR distribution decreases with increasing galaxy density for early-types.  For late-types, the overall SFR distribution is also relatively flat at low densities but then falls to lower values beyond $0.055\,$(Mpc/h)$^{-3}$. However, the 75th percentile of the SFR distribution for late-types does not decrease as sharply as for early-types at these higher densities i.e. the SFR of these high-SF late-type galaxies is higher than for the early-types in the same density regime. Finally, unlike the early-types though, the scatter in the relation is relatively unchanged for late-type galaxies across all densities. 

The results are broadly consistent with the findings of \citet{Gom:03} who also split their sample based on the parameter $C$. They found that early and late-type galaxies each obey a SFR-density relation, although it is a shallow one for early-types (which have low SFR) while for late-types (which dominate the galaxies in their sample with high SFR), the relation is simliar to that for the full sample.

\subsection{Radial variation of SFR as a function of Local Galaxy Density}
\label{subsec:rad_sfr}

Although we have established a correlation between the total galaxy SFR and galaxy density, we can now explore where within the galaxies this suppression is taking place. In particular, we are interested in finding out if the suppression is primarily in the outskirts of galaxies or if it is in their inner regions. The former would support the predictions of hierarchical (``infall and quench'') models of galaxy formation, as the SF in the outskirts of the galaxies would be expected to be first affected by encountering an increasingly dense environment, while the latter, as discussed in \S\,\ref{subsec:evol_vs_env}, would favour an ``environmentally governed evolution'' scenario. By using pixel-z to study the spatial distribution of SF, we will be able to distinguish between these two models.

For each galaxy we calculate a weighted mean SFR $\Psi_{w}$ within
successive annuli:

\begin{equation}
\label{eq:weightedsfr}
\Psi_{w} = \frac{\sum_{i=1}^{N_{\rm a}}w_{i} \times \Psi_{i}} {\sum_{i=1}^{N_{\rm a}}w_{i}}
\end{equation} 
where $w_i$ is the weight corresponding to $\Psi_{i}$, the SFR in pixel
$i$, and $N_{a}$ is the number of pixels within that annulus $a$. The
weight $w_i$ is inversely proportional to the square of the fractional
error on the SFR for the pixel $i$. The SFR in each annulus is thus
normalized by the number of pixels, accounting for the larger total SF at
larger radii due to a greater number of pixels. Scaling the radius of
each annulus by the Petrosian radius ($R_p$) of the galaxy allows us to
stack annuli for different galaxies (which have different Petrosian radii)
together. The center of the galaxy is chosen as the brightest pixel in the galaxy. 

We examine the annular distribution of $\Psi_{w}$ for the stacked
galaxies, spanning a range of local densities:
$0.0 < \rho \le 0.01 \,$(Mpc/h)$^{-3}$,
$0.01 < \rho \le 0.04 \,$(Mpc/h)$^{-3}$,
$0.04 < \rho \le 0.09 \,$(Mpc/h)$^{-3}$.
Figure \ref{fig:NEWRADIALPLOT} shows the
median and 75th percentiles of this distribution. The inner annuli are
more finely binned with $\Delta\,(r/R_p)=0.125$ for $r/R_{p} \le 0.25$,
while $\Delta\,r/R_p=0.25$ for $r/R_{p}>0.25$. Errors are calculated for each annulus by propagating the errors in all the SFRs $\Psi_{i}$ of the pixels within that annulus. The typical size of $0.125 r/R_{p}$ is 2-3 pixels. There are 2785 galaxies where $0.125 r/R_{p}$ is below the pixel resolution. Removing these galaxies neither significantly changes the radial profile of SF nor affect its dependence on local density. 

Star formation in galaxies on average is higher in the the mid-annular region than in the core or outskirts. It is lower (averaged over all morphological types and inclinations) in the nucleus than in the circum-nuclear regions. We find that in dense environments this low SF in the nuclear regions appears lower still. The higher SF in the circum-nuclear region ($0.125 < r/Rp <= 0.25$) is also lessened in more dense regions. It is this depression in the innermost annuli that accounts for the dependence of total galaxy SFR on density.

The effect can be seen most prominently in the 75th percentile of
the distribution of $\Psi_{w}$. The density dependence in the 75th
percentiles of $\Psi_{w}$ is most evident in the first two innermost
annuli up to $r/R_{p}=0.25$,  where there is a clear suppression of
$\Psi_{w}$ between the lowest and highest density intervals: by $8.2 \times 10^{-4} M_{\odot} \, \rm{yr^{-1}}$ (with $4\,\sigma$ significance) in the nuclear region and $2.6 \times 10^{-3} M_{\odot} \, \rm{yr^{-1}}$ (with $3.5\,\sigma$ significance) in the circum-nuclear region. 
Beyond $r/R_{p}=0.25$ there appears to be no clear dependence of the mean SFR
on the local density of galaxies. We see a similar, though less pronounced,
trend in the median of the distribution of $\Psi_{w}$. Again, there is no statistically significant dependence of the mean SFR on local density in galaxy outskirts.

\subsection{Radial variation of SFR with Environment for High and Low Star Forming Galaxies}
\label{subsec:rad_sfr_highandlowsf}

\begin{figure}
\centering
\includegraphics[width=.63\textwidth,height=0.46\textwidth]{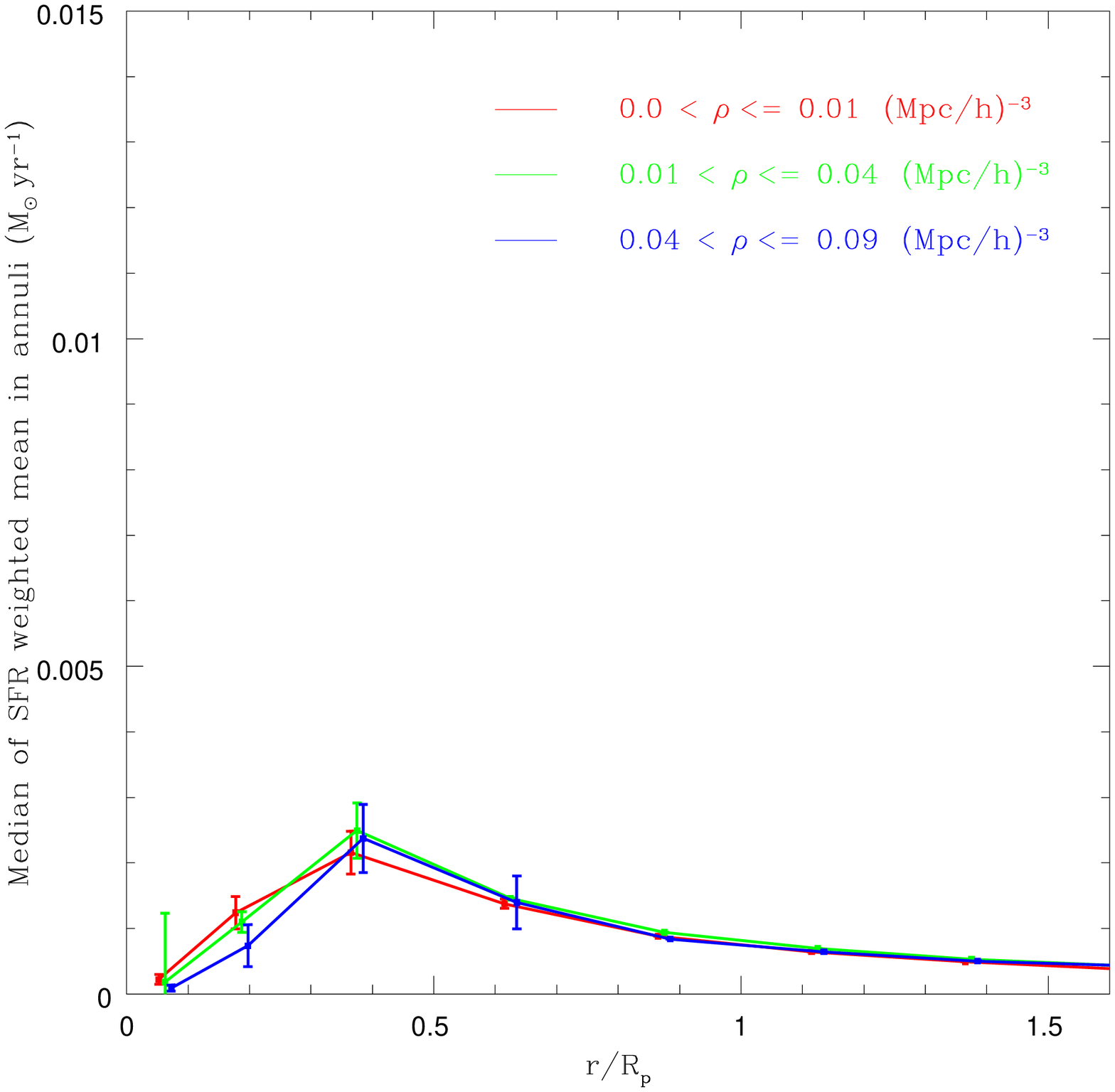}
\includegraphics[width=.63\textwidth,height=0.46\textwidth]{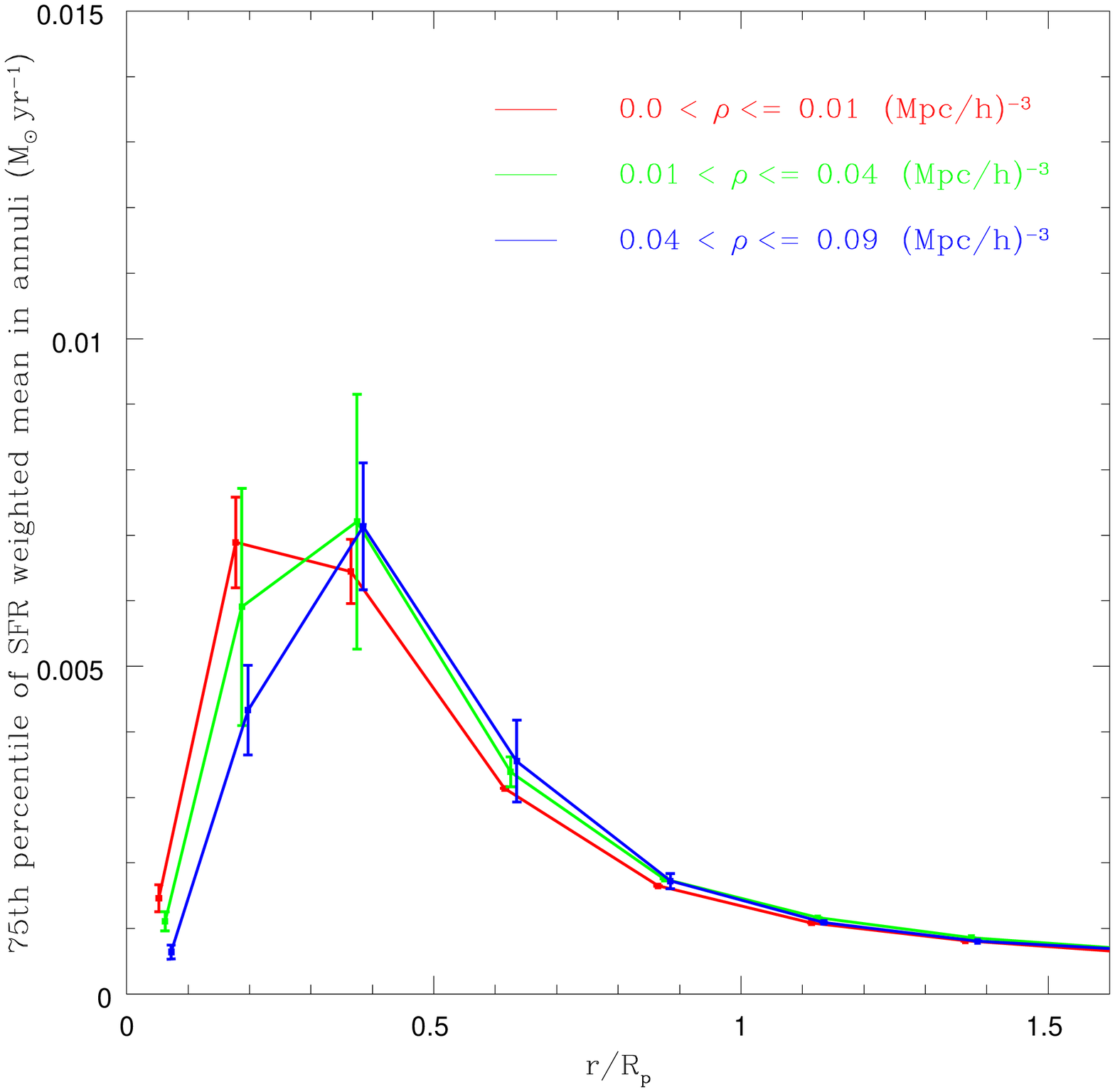}
\caption{Top panel: Median of the weighted mean SFR $\Psi_{w}$ ($M_{\odot} \,yr^{-1} $) within successive radial annuli for 3 different intervals of galaxy density $\rho$: $0-0.01 \,$(Mpc/h)$^{-3}$ (red), $0.01-0.04 \,$(Mpc/h)$^{-3}$ (green), $0.04-0.09 \,$(Mpc/h)$^{-3}$ (blue). Bottom panel: 75th percentile of $\Psi_{w}$ ($M_{\odot} \,yr^{-1} $). The mean SFR in each annulus for each galaxy is a weighted mean of the SFRs in all the pixels in the annulus. For galaxies in each local density interval, these radial annuli are then stacked. The percentiles are obtained from the distribution of the mean SFR in these stacked annuli.  The inner annuli are more finely binned with $\Delta\,r/R_p=0.125$, while $\Delta\,r/R_p=0.25$ for $r/R_{p}>0.25$. The density dependence in the 75th percentiles of $\Psi_{w}$ is most evident in the first two innermost annuli up to $r/R_{p}=0.25$,  where there is a clear suppression of $\Psi_{w}$ between the lowest and highest density intervals. The same is true to a lesser extent in the median. Beyond $r/R_{p}=0.25$, no dependence on the local density of galaxies is detected.  
\label{fig:NEWRADIALPLOT}}
\end{figure}

\begin{figure}
\centering
\includegraphics[width=.63\textwidth,height=0.55\textwidth]{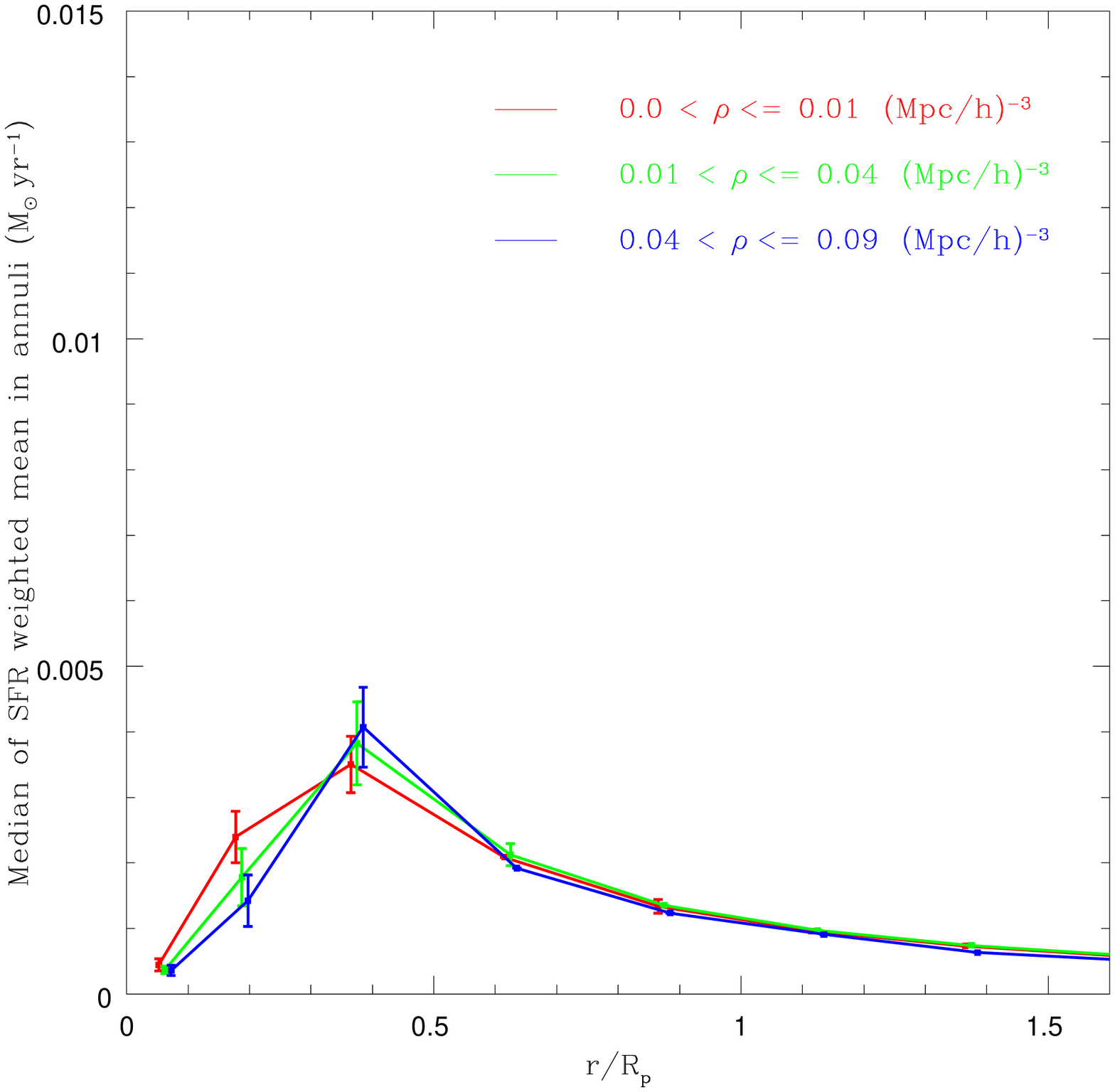}
\includegraphics[width=.63\textwidth,height=0.55\textwidth]{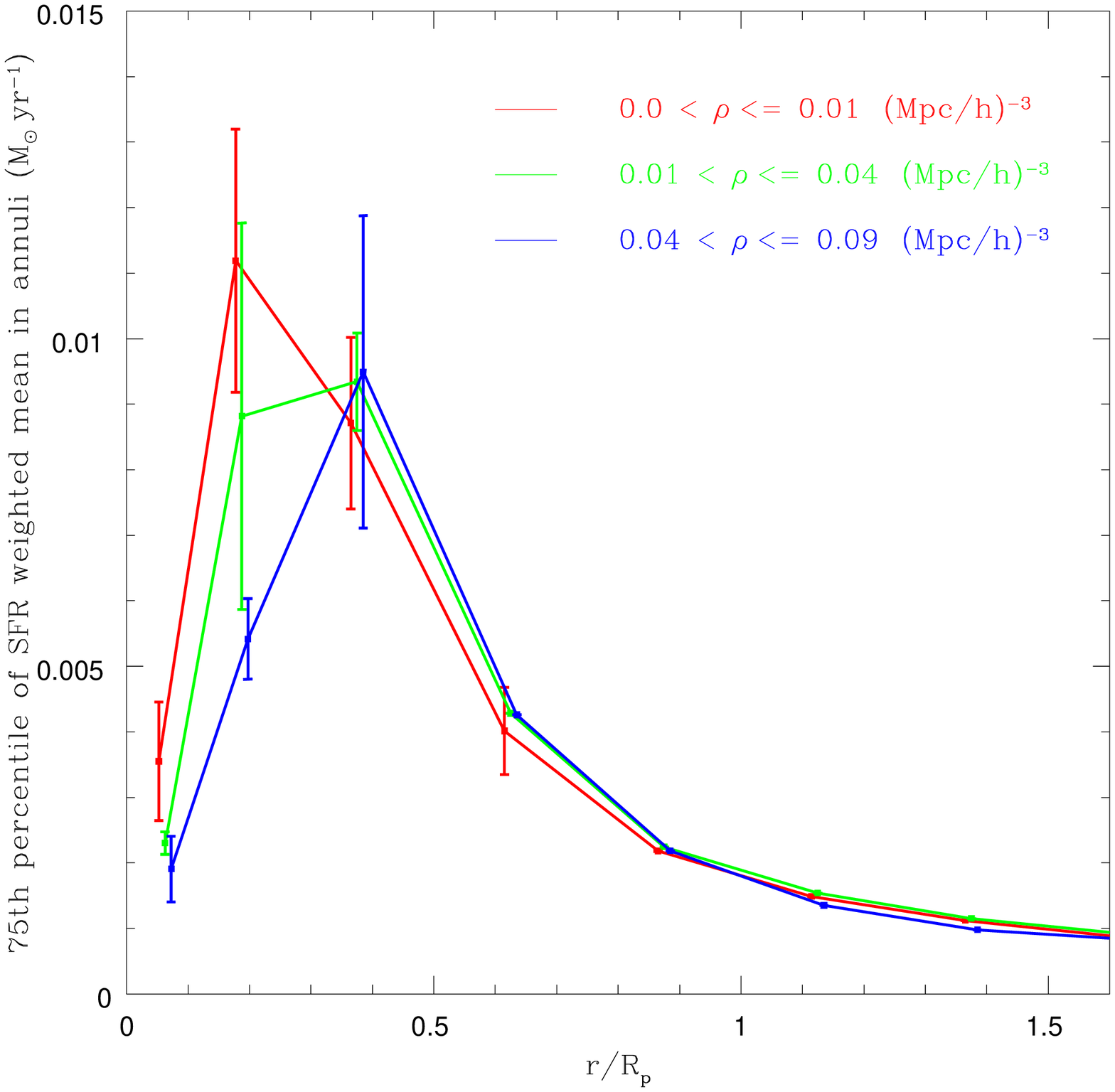}
\caption{Top panel: Median of the distribution of weighted mean SFRs $\Psi_{w}$ ($M_{\odot} \,yr^{-1} $) within successive radial annuli as a function of the local galaxy density $\rho$ for high star forming galaxies ($> 1.02\,M_{\odot} \,yr^{-1}$ ). There are 3 different intervals of $\rho$, as in Figure~\ref{fig:NEWRADIALPLOT}. Bottom panel: 75th percentile of $\Psi_{w}$ ($M_{\odot} \,yr^{-1} $).
\label{fig:NEWRADIALPLOT_HIGHSF}}
\end{figure}

\begin{figure}
\centering
\includegraphics[width=.63\textwidth,height=0.55\textwidth]{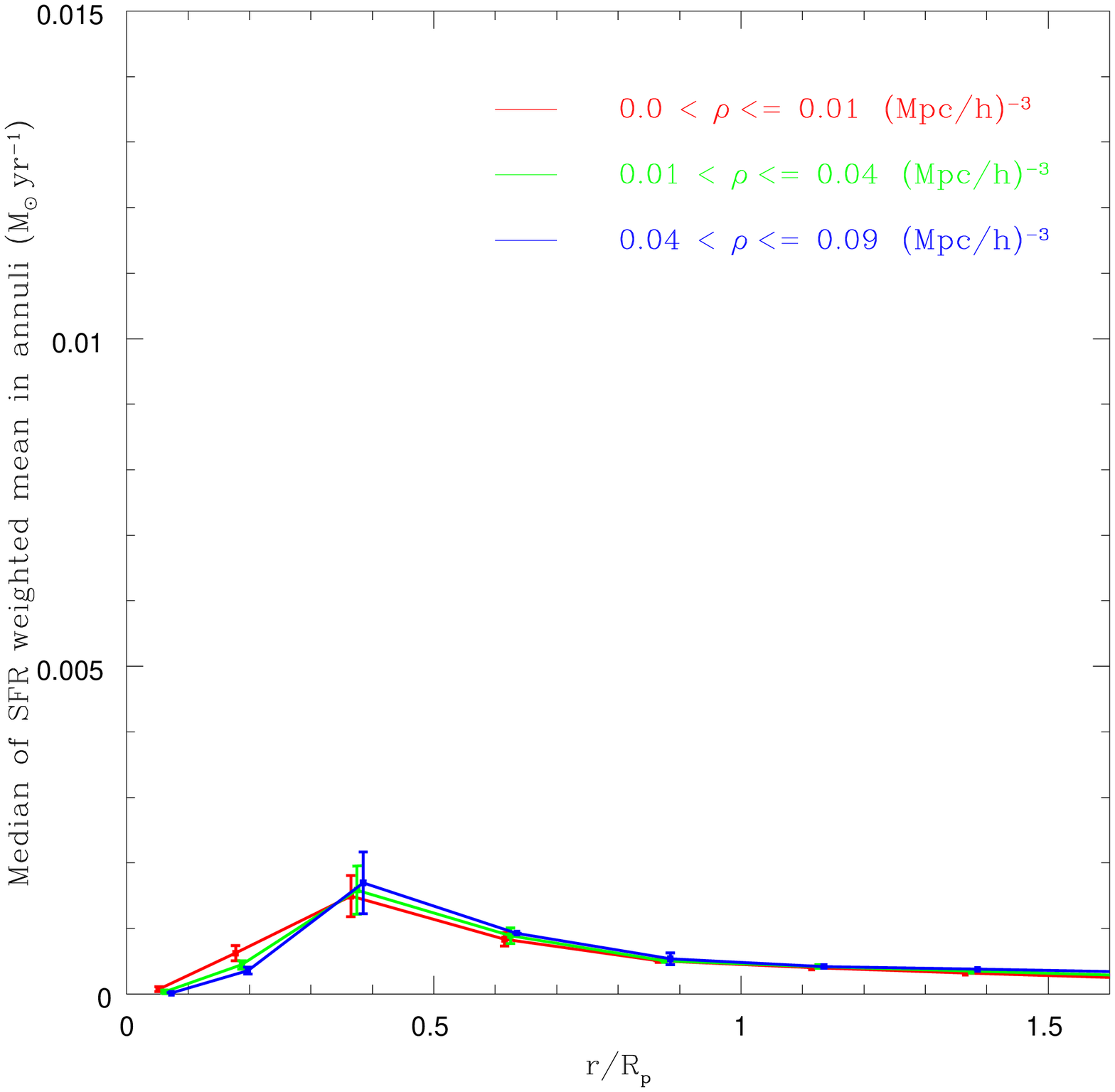}
\includegraphics[width=.63\textwidth,height=0.55\textwidth]{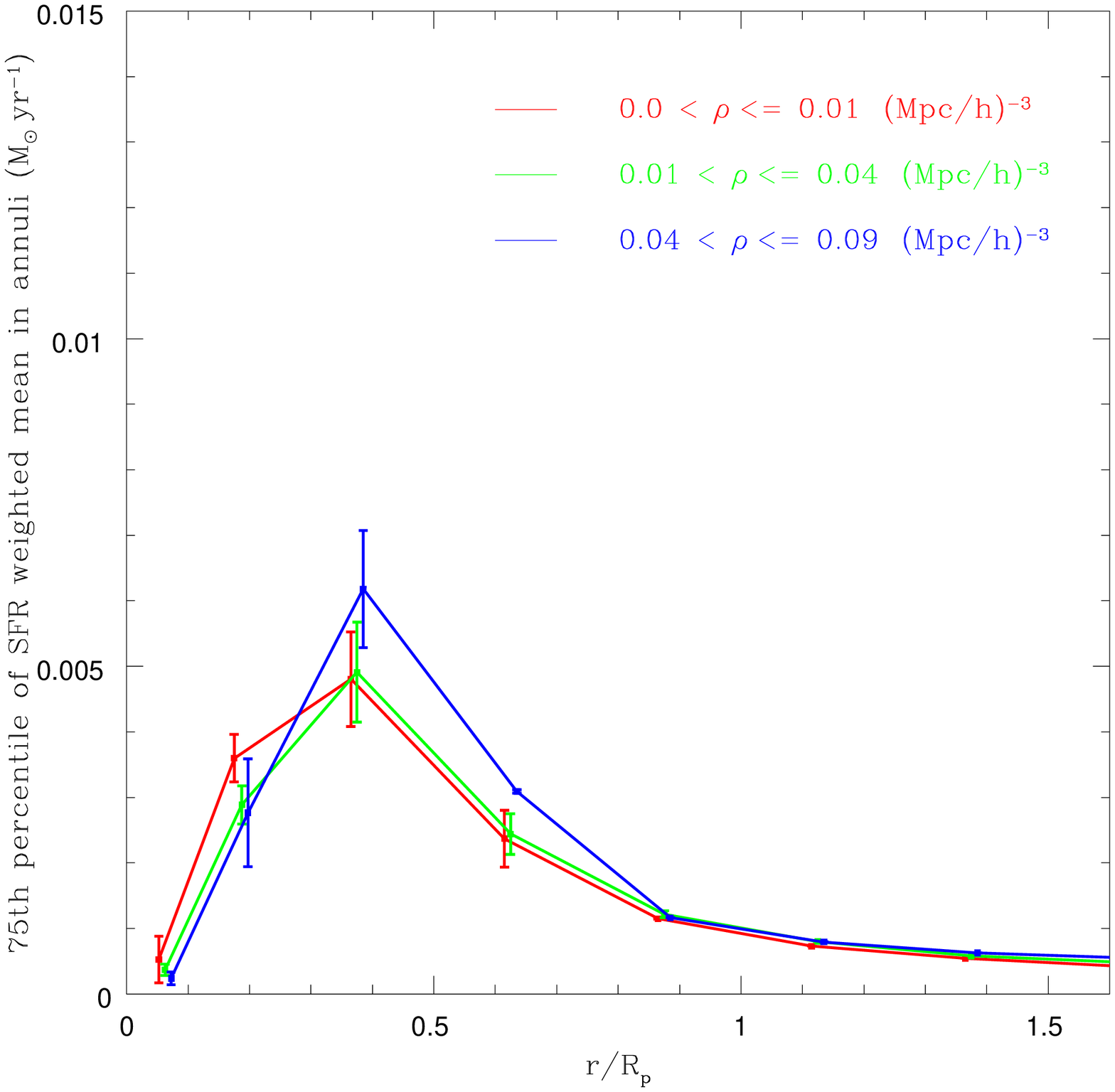}
\caption{Top panel: Median of the distribution of weighted mean SFRs $\Psi_{w}$ ($M_{\odot} \,yr^{-1} $) within successive radial annuli as a function of the local galaxy density $\rho$ for low star forming galaxies ($<0.32\,M_{\odot} \,yr^{-1}$ ). There are 3 different intervals of $\rho$, as in Figure~\ref{fig:NEWRADIALPLOT}. Bottom panel: 75th percentile of $\Psi_{w}$ ($M_{\odot} \,yr^{-1} $).
\label{fig:NEWRADIALPLOT_LOWSF}}
\end{figure}

The suppression of the total galaxy SFR with increasing local density is
most noticeable in the 75th percentile curve of Figures~\ref{fig:totalsfr}.
In other words, at progressively higher densities, the distribution
of galaxy SFRs is truncated at lower SFR values, and it is the population of
the most highly star forming galaxies that are being affected.
Here we investigate whether the radial variation we observe in
Figure \ref{fig:NEWRADIALPLOT} is seen only in these high SFR galaxies,
or whether such a trend also exists in more quiescent galaxies. We
examine the galaxy populations comprising the upper and lower quartile
of the SFR distribution, and the results are shown in
Figure~\ref{fig:NEWRADIALPLOT_HIGHSF} for the high SFR galaxies
(with $\rm{SFR}>1.02 \,M_{\odot} \, \rm{yr^{-1}} $) and in
Figure~\ref{fig:NEWRADIALPLOT_LOWSF} for the low SFR galaxies
($\rm{SFR}<0.32 \,M_{\odot} \, \rm{yr^{-1}} $). For the high SFR population,
up to a radius of $0.25 R_p$, the suppression of mean SFR with environment
is clearly evident, and as with the total galaxy population, there is no dependence
beyond this radius. Again, as before, the effect is most pronounced for the 75th percentile of the mean SFR (in the range $0 < r/Rp \le 0.25$) , where there is a $2\,\sigma$ difference between the highest and lowest density intervals in the range $0 < r/Rp \le 0.125$ and a $3\,\sigma$ difference in the range $0.125 < r/Rp \le 0.25$. But it can also be seen in the median of the distribution (in the range $0.125 < r/Rp \le 0.25$), where the difference is $2.5\,\sigma$. This is consistent with the dependence of the radial variation of the mean annular SF with environment for the total galaxy population in the sample (averaged over all star formation rates) in \S\,\ref{subsec:rad_sfr}.

For the low star forming galaxies a small, although not statistically
significant, dependence on environment is seen. It is thus the population
of strongly star forming systems that account for the observed relation
between the radial mean SFR and local density of galaxies shown
in Figure \ref{fig:NEWRADIALPLOT}. The suppression of SFR in the innermost
annuli of the most strongly star-forming systems accounts for
the observed SFR-density relation in the overall galaxy population.

\subsection{Physical Interpretation}
\label{subsec:interpretation}

In \S\,\ref{subsec:totsfr} and \S\,\ref{subsec:totsfr_types}, we find evidence for a suppression of total SFR as we go to higher density environments, and this relation seems to be preserved independent of morphology. So the total SFR-density relation is a result of SFR suppresion at higher densities, not just a result of the morphology-density relation (at least not solely). The suppression is evident at the highest densities, beyond $0.05\,$(Mpc/h)$^{-3}$, in the regime of clusters. In these high density environments, there are a number of physical mechanisms responsible for the suppression of total SFR. These include ram-pressure stripping, galaxy harassment, and tidal disruptions, mechanisms that are known to be dominant in cluster cores. 

However, in \S\,\ref{subsec:rad_sfr}, we find that the suppression of mean radial SF takes place in the cores of galaxies (particularly in the nuclear and circum-nuclear region) while the outskirts are not affected by their changing environment. This implies that the environment itself is not impinging on galaxies to suppress the SFR, as the outer regions should otherwise be most affected according to ``infall and quench'' models.  Any physical mechanism cannot therefore be solely responsible for this drop in the nuclear SF. By extension, they cannot therefore be governing the suppression in total SFR in galaxies in high density environments.  

This seems to point to an evolutionary rather than an environmental origin for the SFR-density relation. 
If galaxies in more dense environments formed their stars earlier and faster than those in
less dense environments, that would be consistent with either a uniformly
depressed SFR as a function of galaxy radius, or a depressed nuclear SFR,
given that higher gas densities in the nuclear regions are likely to
drive star formation more rapidly (supported by the fact that star
forming disk galaxies typically show bulges dominated by old, red stellar
populations).

These results seem to be consistent with the idea of ``downsizing''
in galaxy formation \citep{Cow:96}, whereby the more massive galaxies form
at earlier epochs. Downsizing is characterised by a decline in the mass of
the galaxies that dominate the star-formation rate density with
decreasing redshift (``downsizing of star formation''). This is supported by recent measurements of star-formation histories in both local galaxies from the SDSS \citep{Hea:04} and distant galaxies from the Gemini Deep Deep survey \citep{Jun:05}.
This is distinct from the ``downsizing with quenching'' which follows a different timescale - spheroidal galaxies have been known to have a second star-formation timescale, namely that of ``quenching'': it becomes easier to keep galaxies gas-free with time. In fact, recent studies of the galaxy luminosity function
at $z\approx 1$ \citep{Bel:04,Fab:05} conclude that massive red galaxies observed at $z\approx 0$ migrated to the bright
end of the red sequence by a combination of two process: the quenching of
star formation in blue galaxies and the merging of less-luminous, previously quenched red galaxies. \citet{Fab:05}
conclude that the typical mass at which a blue, star-forming galaxy is
quenched (and therefore enters the red sequence) decreases with time. Our radial result is evidence for the ``downsizing of star formation'' as opposed to this ``downsizing with quenching''. 

In \S\,\ref{subsec:rad_sfr_highandlowsf}, we have also found that whereas the quiescent galaxies (corresponding to the bottom quartile of the total SFR distribution) in the sample show no significant dependence of SFR on environment at any radius, the radial dependence of the mean annular SFR for the highest star-forming galaxies (top quartile of the total SFR distribution) mirrors the trend for the full sample. In addition, while the decrease of total SFR with density is dominated by the high-SFR galaxies, the quiescent galaxies show hardly any dependence of their total SFR on environment. The radial result in \S\,\ref{subsec:rad_sfr_highandlowsf} confirms that the total SFR-density relation is largely due to the decline of nuclear SF in these high SF galaxies as we go to higher density environments. The overall radial result can therefore be explained if we consider a population of active, star-forming galaxies which formed at high redshift. The SF in this population of galaxies is dominated by the nuclear SF. Out of this population, the more massive systems which formed in more dense environments also formed their stars earlier and faster than those in less dense environments. By $z\approx 0$, this sub-population (which originated from more massive systems) has a lower nuclear SF than the sub-population which formed in less dense environments. This supports the idea that the ``downsizing of star formation'' in galaxies as described by \citet{Cow:96} apples only to the actively star-forming galaxies in our sample. 

It could also be possible that this SFR suppression in galaxy cores may
be a consequence of feedback from an active galactic nucleus (AGN).
In particular, if AGN feedback in more massive systems is more efficient than
in lower mass systems, and since dense environments are known to host more
massive galaxies on average \citep{Dre:80,Pos:84} this could perhaps lead to
preferential suppression of nuclear star formation in more dense
environments. This mechanism could be explored further by looking into the
mass dependence of the SFR suppression in galaxies, although this is beyond
the scope of our current analysis. The radius out to which such a feedback mechanism
could suppress SF also needs to be quantified. The complete explanation is expected to be a combination of this ``downsizing'' in SF together with more efficient AGN feedback in galaxies in more dense environments.

\subsection{Future Work}
\label{subsec:future}

Although the density-morphology relation is seen to be independent of the total SFR-density relation, the relationship between the density-morphology relation and the radial result has yet to be explored. We know that in addition to the nuclear SF declining in more dense environments, the location of the SF peak within galaxies is shifted to larger radii at higher densities. Is this telling us that the signal from early-type galaxies is just becoming stronger as we go to higher densities because there is a higher proportion of early types in those environments? At lower densities, the nuclear SF is higher, so could this be due to a predominance of late-types which are known to form in these low density environments? We know that by the ``downsizing'' argument, the more massive galaxies at high redshift form their stars early and rapidly. But these more massive systems tend also to to be the elliptical galaxies, so to some degree, the radial trend (and its dependence on density) for the full sample could still be, at least in part, a result of the density-morphology relation.  To disentangle this, we wish to find out first what the radial SFR-density trends look like for each morphological type within our sample. Next, given these separate radial profiles, together with the fact that the proportion of early-type galaxies is expected to increase as we go to intervals of higher density, we wish to establish what the contribution of each type is to the radial SF-density trend for the full sample. 

As discussed above in \S\ref{subsec:rad_sfr_highandlowsf}, we also know that it is indeed a sub-population of active, star-forming galaxies that is responsible for the trend of radial SF with environment. We wish to establish whether these systems are dominated by late-type galaxies or if they are a mixture of early and late-types. If this sub-population is indeed primarily late-type galaxies, then the radial SFR-density trend for late-type galaxies should mirror that of the high-SF population.

To further explore and support the ``downsizing'' idea, we also plan to extend this work to include an analysis on galaxies in a high redshift sample, from
COSMOS \citep{Koe:07},
GOODS \citep{Gia:04}, or
AEGIS \citep{Dav:07}. At high redshifts the high density
environments are less evolved, and the relationship between SFR and
density is likely to be quite different. Indeed, using GOODS data,
\citet{Elb:07} has shown that the SFR-density
relationship is inverted by $z\approx 1$, such that dense environments
support enhanced, rather than suppressed, star formation.
Using DEEP2 Galaxy Redshift Survey data, \citet{Coo:06} also suggest
that bright blue galaxies in overdense regions at high redshift have their
star formation quenched and evolve into members of the red sequence by
$z\approx 0$. These conclusions indicate that there exists a population
of massive blue galaxies that has likely undergone quenching at $z\approx 1$,
which has no high-mass counterpart today, and that there is a downsizing
of the characteristic mass (or luminosity) at which the quenching
of a galaxy's star formation becomes efficient. They
suggest that the quenching mechanism must operate efficiently in both cluster
and group environments for consistency between their $z\approx 1$ and
$z\approx 0$ results. Using pixel-z to explore the relationship between
environment and the SFR distribution within galaxies at high redshift will
aid in constraining the process by which such a mechanism might operate.

Lastly, we aim to explore the differences that may arise from using templates from a different stellar population synthesis model, in order to determine if the statistical results are in fact dependent on the model chosen and the different degeneracies that exist among the various parameters in that model. In future work, we aim to explore this issue using different models, such as PEGASE \citep{Fio:97}.

\section{Conclusions
\label{sec:conclusions}}

Pixel-z is a useful technique for exploring spatially
distributed galaxy properties, including the star formation rate. Although it has inherent limitations (detailed in this paper and by \citet{Con:03}), it can provide insights to
direct more detailed exploration, and when applied to large samples, many of those limitations can be overcome. We draw several conclusions about the environmental dependence of spatially resolved galaxy star formation:

\begin{enumerate}
\item By summing the SFRs in individual pixels, a total SFR-density relation is established. The SFR-density relation measured from emission lines (\citet{Gom:03}, \citet{Lew:02}) is confirmed based on SFRs inferred from SED template-fitting to pixel fluxes across five bands.
\item The total SFR-density relation is found to be independent of morphological type of the galaxy. Therefore, the SFR-density relation of galaxies in the SDSS is not solely a result of having a larger fraction of early types as we go to higher densities (the morphology-density relation). So there is a SFR suppression in more dense environments. 
\item Within galaxies, star formation, averaged over all morphological types and
inclinations, is highest in a region spanning $0.125\lesssim r/R_p \lesssim 0.5$, with the specific location of the
peak depending on local density. It is lower in both the nucleus and
the outskirts than in the circum-nuclear region. In denser environments,
the low mean SFR in the nuclear regions appears lower still. The
higher mean SFR in the circum-nuclear region is also lessened in
more dense regions, with the peak SFR moving towards larger radii. The SFR beyond $r/R_p\approx 0.25$ and in the outskirts of galaxies is not affected by a changing environment. It is thus a depression of SF in the innermost annuli that accounts for the dependence of total galaxy SFR on density. 

\item When the sample is split based on the total SFR of galaxies, it is the highly star-forming systems that are found to be responsible for the dependence of radial SFR on local density. The low star-forming galaxies show little dependence of radial SFR on environment.
\end{enumerate}

\section{Acknowledgements
\label{sec:acknowledge}}

NW would like to thank Tam{\'a}s Budav{\'a}ri, Ravi K. Sheth, Ching-Wa Yip, Jeffrey Gardner, Simon Krughoff, Samuel Schmidt, Cameron McBride and Jeremy Brewer for useful discussions. AJC and NW acknowledge partial support from NSF grant ITR- 0312498 and NASA AIRSP award NNG06GH97G. AMH acknowledges support provided by the Australian Research Council in the form of a QEII Fellowship (DP0557850). This material is based upon work supported by the National Science Foundation under the following NSF programs: Partnerships for Advanced Computational Infrastructure, Distributed Terascale Facility (DTF) and Terascale Extensions: Enhancements to the Extensible Terascale Facility.

Funding for  the SDSS and SDSS-II  has been provided by  the Alfred P.
Sloan Foundation, the Participating Institutions, the National Science
Foundation, the  U.S.  Department of Energy,  the National Aeronautics
and Space Administration, the  Japanese Monbukagakusho, the Max Planck
Society, and  the Higher Education  Funding Council for  England.  The
SDSS Web  Site is  http://www.sdss.org/.  The SDSS  is managed  by the
Astrophysical    Research    Consortium    for    the    Participating
Institutions. The  Participating Institutions are  the American Museum
of  Natural History,  Astrophysical Institute  Potsdam,  University of
Basel,   Cambridge  University,   Case  Western   Reserve  University,
University of Chicago, Drexel  University, Fermilab, the Institute for
Advanced   Study,  the  Japan   Participation  Group,   Johns  Hopkins
University, the  Joint Institute  for Nuclear Astrophysics,  the Kavli
Institute  for   Particle  Astrophysics  and   Cosmology,  the  Korean
Scientist Group, the Chinese  Academy of Sciences (LAMOST), Los Alamos
National  Laboratory, the  Max-Planck-Institute for  Astronomy (MPIA),
the  Max-Planck-Institute  for Astrophysics  (MPA),  New Mexico  State
University,   Ohio  State   University,   University  of   Pittsburgh,
University  of  Portsmouth, Princeton  University,  the United  States
Naval Observatory, and the University of Washington.

\acknowledgments

\end{document}